\documentclass[12pt,letterpaper]{article}

\usepackage{natbib,hyperref}
\hypersetup{colorlinks=true, citecolor=blue!50!black, linkcolor=red!60!black, urlcolor=blue!50!black}

\makeatletter
\g@addto@macro\@floatboxreset{\def\baselinestretch{1}\normalsize}
\makeatother

\newcommand{\blind}{0}
\newcommand{\spacingset}[1]{\renewcommand{\baselinestretch}%
{#1}\small\normalsize}
\usepackage{graphicx,bm,colonequals,amsmath,amssymb,url,xcolor,bbm}
\usepackage{array,tabularx,multirow,booktabs}
\usepackage[font={footnotesize}]{caption, subcaption}
\usepackage[utf8]{inputenc}
\usepackage{enumitem}
\usepackage{soul} 
\usepackage{placeins} 
\usepackage[normalem]{ulem}
\usepackage{amsmath,amssymb,amsthm}

\graphicspath{{figures/}}
\newcommand{\figureorplaceholder}[2]{\IfFileExists{figures/#2}{\includegraphics[#1]{#2}}{\includegraphics[#1]{example-image}}}

\newcommand{\NNV}{NNV}
\newcommand{\cs}{m}

\usepackage{xr}
\makeatletter
\newcommand*{\addFileDependency}[1]{
  \typeout{(#1)}
  \@addtofilelist{#1}
  \IfFileExists{#1}{}{\typeout{No file #1.}}
}
\makeatother

\usepackage{mathtools}
\mathtoolsset{showonlyrefs} 

\everydisplay{\textstyle}

\bibpunct[, ]{(}{)}{;}{a}{,}{,}
   
\usepackage{amsthm}
\newtheoremstyle{propstyle} 
    {2mm}                    
    {1mm}                    
    {\itshape}                   
    {}                           
    {\scshape}                   
    {.}                          
    {.5em}                       
    {}  
\theoremstyle{propstyle}
\newtheorem{proposition}{Proposition}
\theoremstyle{propstyle}

\theoremstyle{propstyle}

\theoremstyle{propstyle}

\theoremstyle{propstyle}

\usepackage[ruled,vlined]{algorithm2e}
\SetKwInput{KwInput}{Input}
\usepackage{algorithmic}
\usepackage{array,framed}
\usepackage{float}

\makeatletter
\renewcommand{\paragraph}{%
  \@startsection{paragraph}{4}%
  {\z@}{2ex \@plus 1ex \@minus .2ex}{-1em}%
  {\normalfont\normalsize\bfseries}%
}
\makeatother

\DeclareMathAlphabet\mathbfcal{OMS}{cmsy}{b}{n}

\renewcommand{\bf}{\mathbf{f}}
\newcommand{\bh}{\mathbf{h}}
\newcommand{\bu}{\mathbf{u}}

\newcommand{\bs}{\mathbf{s}}

\newcommand{\bx}{\mathbf{x}}
\newcommand{\by}{\mathbf{y}}

\newcommand{\bw}{\mathbf{w}}
\newcommand{\bS}{\mathbf{S}}

\newcommand{\bz}{\mathbf{z}}

\newcommand{\bA}{\mathbf{A}}

\newcommand{\bW}{\mathbf{W}}
\newcommand{\bF}{\mathbf{F}}
\newcommand{\bG}{\mathbf{G}}

\newcommand{\bL}{\mathbf{L}}

\newcommand{\bI}{\mathbf{I}}
\newcommand{\bD}{\mathbf{D}}
\newcommand{\bH}{\mathbf{H}}
\newcommand{\bU}{\mathbf{U}}
\newcommand{\bV}{\mathbf{V}}

\newcommand{\bQ}{\mathbf{Q}}
\newcommand{\bB}{\mathbf{B}}
\newcommand{\bC}{\mathbf{C}}

\newcommand{\bfzero}{\mathbf{0}}

\newcommand{\bfmu}{\bm{\mu}}

\newcommand{\bftheta}{\bm{\theta}}

\newcommand{\bfnu}{\bm{\nu}}

\newcommand{\bfSigma}{\bm{\Sigma}}

\newcommand{\order}{\mathcal{O}}
\newcommand{\normal}{\mathcal{N}}

\DeclareMathOperator*{\argmin}{arg\,min}
\DeclareMathOperator*{\KL}{KL}

\newcommand{\domain}{\mathcal{D}}

\theoremstyle{remark}
\newtheorem{remark}{Remark}
\newcommand{\Srho}{\mathbf{S}_{\rho}}       
\newcommand{\SA}{\mathbf{S}^{L}_{\rho}}     
\newcommand{\tW}{\widetilde{\mathbf{W}}}
\newcommand{\hU}{\widehat{\mathbf{U}}}
\newcommand{\hL}{\widehat{\mathbf{L}}}
\newcommand{\cK}{\mathcal{K}}           
\newcommand{\cJ}{\mathcal{J}}           
\newcommand{\bt}{\mathbf{t}}

\begin{document}

\def\spacingset#1{\renewcommand{\baselinestretch}%
{#1}\small\normalsize} \spacingset{1}


\if0\blind
{
  \title{\vspace{-1.2cm}\bfseries Nearest-Neighbor Sparse Cholesky Decomposition with Applications to High-Dimensional Spatio-Temporal Filtering}
  \author{Pratik Nag\hspace{.2cm}\\
    Department of Statistics, University of Wisconsin--Madison\\
    and \\
    Matthias Katzfuss\thanks{Corresponding author: \texttt{katzfuss@gmail.com}} \\
    Department of Statistics, University of Wisconsin--Madison}
  \date{}
  \maketitle
  \vspace{-1.2cm}
} \fi

\if1\blind
{
  \bigskip
  \bigskip
  \bigskip
  \begin{center}
    {\LARGE\bfseries Nearest-Neighbor Sparse Cholesky Decomposition with Applications to High-Dimensional Spatio-Temporal Filtering}
  \end{center}
  \medskip
} \fi

\begin{abstract}
Filtering for high-dimensional spatio-temporal state-space models requires repeated operations with posterior covariance matrices, which is infeasible for exact Kalman filters on fine spatial grids. Nearest-neighbor Vecchia approximations are among the most accurate scalable approximations for spatial inference, but they yield sparse precision factors, whereas the forecast step of a filter requires covariances. We propose a three-stage sparse Cholesky decomposition that turns this conditional sparsity into generative sparsity. A forward-KL-optimal Vecchia factor approximates the prior precision, incomplete Cholesky factorizes the resulting sparse posterior precision, and a sparse approximate inverse recovers a sparse posterior covariance factor. The last stage decouples into small independent least-squares problems with closed-form solutions, and its objective controls the reverse KL divergence. The covariance factor gives forecast covariances and samples without any global linear solve, at quasilinear cost and in parallel; beyond filtering, it provides posterior variances and conditional simulations for other latent Gaussian models. Because its nearest-neighbor conditioning sets do not depend on a partition of the domain, the filter is free of the artifacts along subregion boundaries that hierarchical Vecchia (HV) filters produce in simulated and real data. Non-Gaussian data are handled by a Laplace approximation. In simulations with Gaussian, logistic, Poisson, and gamma data, the proposed filter is close to the exact filter, and at the same sparsity of the posterior covariance factor, its loss in log score relative to the exact filter is about $8$ to $28$ times smaller than that of HV and $100$ to $350$ times smaller than that of low-rank filters, while it is at least as fast as both (for $90{,}000$ locations, about twice as fast for Gaussian data and three to five times as fast for non-Gaussian data). In a filtering analysis of satellite measurements of total precipitable water vapor on a grid of $61{,}009$ cells (chosen so that the exact Kalman filter is just about feasible for comparison), the proposed filter is essentially indistinguishable from the exact filter at held-out observations, with a root mean squared prediction error only $0.1\%$ above that of the exact filter, compared with $4.5\%$ for HV and $159\%$ for the low-rank filter, and it is about three times as fast as HV and almost twice as fast as the low-rank filter.
\end{abstract}

\noindent%
\vfill



\section{Introduction \label{sec:intro}}

Spatio-temporal state-space models, in which a latent Gaussian field evolves over time and is observed through noisy and often non-Gaussian measurements, are central to the analysis of remote-sensing data and to data assimilation \citep{Mardia1998,Wikle1999,Cressie2011,Wikle2010a}. For linear Gaussian models, the Kalman filter \citep{Kalman1960,Anderson1979} gives the exact filtering distributions, but each of its steps requires operations on dense $n\times n$ covariance matrices and therefore costs $\order(n^3)$. For fine spatial grids, with $n$ in the tens of thousands or more, the exact filter is infeasible and must be approximated.

Many approximations have been proposed. Ensemble Kalman filters \citep{Evensen1994,Burgers1998,Houtekamer2016,Katzfuss2015b} are the workhorse of operational data assimilation but incur sampling error that requires localization and tuning, and particle filters break down in high dimensions \citep{Snyder2008}. Low-rank filters \citep{Verlaan1995,Wikle1999,Cressie2010a} are deterministic and fast but cannot capture fine-scale variation \citep{Stein2013a}. Models based on stochastic partial differential equations (SPDEs) have sparse precision matrices via Gaussian Markov random field approximations \citep{Lindgren2011a,Cameletti2013}, whose sparsity is not preserved by filtering, and spectral methods give fast filters for advection--diffusion SPDEs with constant coefficients on regular grids \citep{Sigrist2015}. Multi-resolution filters \citep{Jurek2018} and the hierarchical Vecchia (HV) filter \citep{Jurek2020} overcome this limitation with block-sparse hierarchical decompositions that are preserved through the forecast and update steps.

Among the most accurate scalable methods for spatial inference are Vecchia approximations \citep{Vecchia1988,Stein2004,Datta2016,Guinness2016a,Katzfuss2017a}, which can be viewed as sparse inverse Cholesky factorizations of the covariance matrix. Under a maximin ordering with nearest-neighbor conditioning, the factor that minimizes the forward Kullback--Leibler (KL) divergence is available in closed form \citep{Schafer2020}. Because of the screening effect \citep{Stein2002}, it is highly accurate in numerical studies \citep[e.g.,][]{Guinness2016a,Katzfuss2017a}, with theoretical guarantees for certain Mat\'ern-type covariances \citep{Schafer2020,Kang2024}. For latent Gaussian processes, this sparse prior precision can be combined with incomplete Cholesky factorization \citep{Schafer2020}, Laplace approximations \citep{Zilber2019}, iterative solvers \citep{Kundig2024}, or variational inference; \citet{Cao2023} minimize a forward KL for the prior and a reverse KL for the posterior over sparse inverse Cholesky factors.

These precision-based methods do not directly carry over to filtering. The forecast step propagates the posterior \emph{covariance} matrix, which a sparse precision factor provides only through global linear solves. HV circumvents this problem by restricting the conditioning sets to a block-hierarchical domain partitioning, under which the precision and covariance factors share the same sparsity and the required factorizations and inversions are exact \citep{Jurek2020}. However, hierarchical conditioning sets approximate the prior far less accurately than nearest-neighbor sets of the same size \citep{Katzfuss2017a}, and the partitioning causes artifacts along subregion boundaries \citep[][Secs.~3.4 and~S5]{Jurek2020}.

\begin{figure}[t!]
    \centering
    \includegraphics[width=0.78\textwidth]{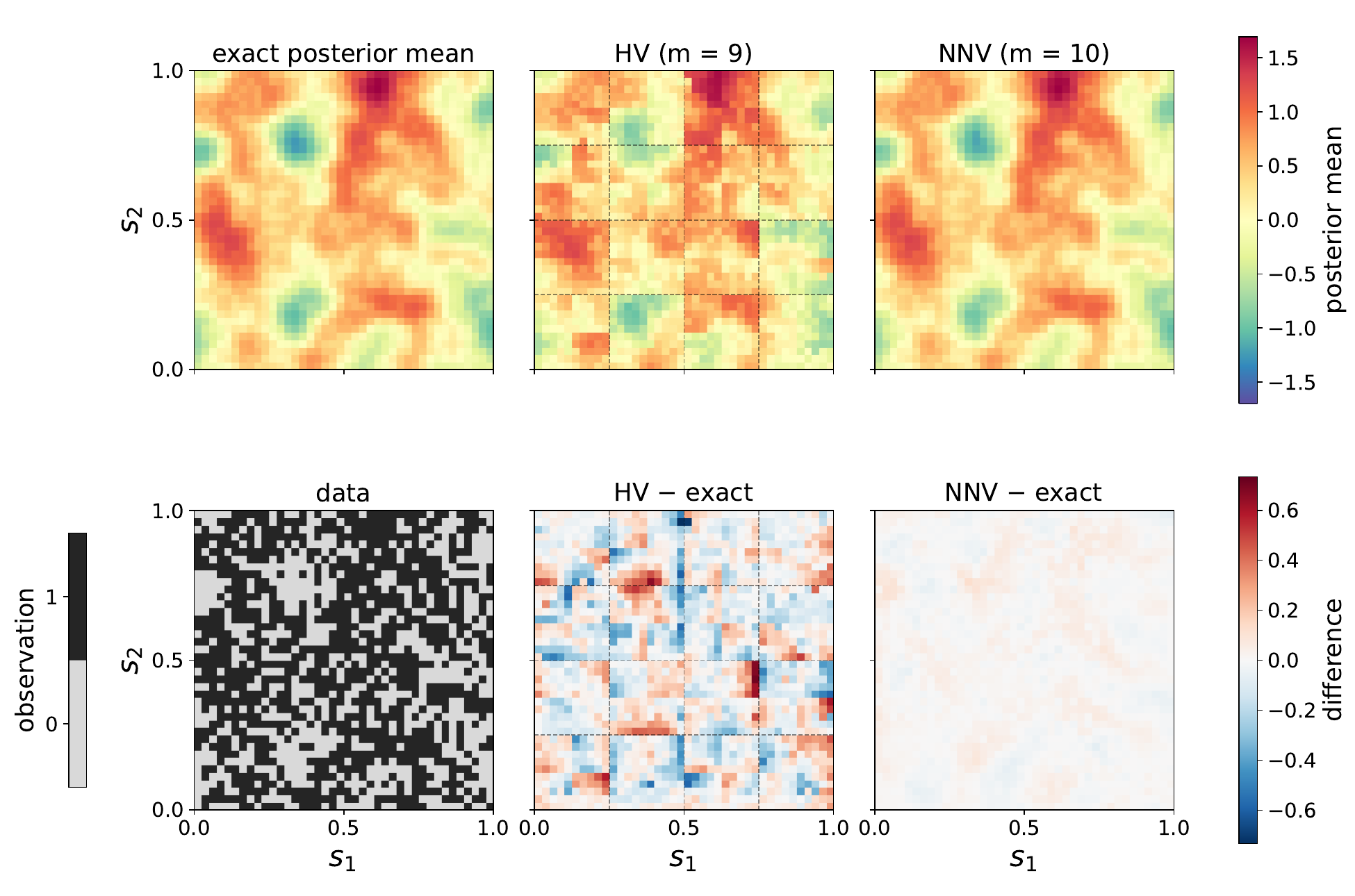}
    \caption{Simulated binary observations at $n=1{,}600$ locations with a logistic link (bottom left; Section~\ref{sec:boundary}). Top row: posterior means of the latent field using the exact Laplace approximation, HV ($\cs{}=9$), and \NNV{} ($\cs{}=10$), on a common color scale, where $\cs$ is the median conditioning-set size (Section~\ref{sec:setup}). Bottom center and right: differences of the HV and \NNV{} posterior means from the exact one, on a common color scale. The HV posterior mean has artifacts along the boundaries of its domain partition, most visibly along $s_1=1/2$, whereas \NNV{} is visually indistinguishable from the exact posterior mean. Dashed lines in the HV panels mark the coarse partition lines of HV at $s_1,s_2\in\{1/4,1/2,3/4\}$.}
    \label{fig:boundary}
\end{figure}

We propose a nearest-neighbor sparse Cholesky decomposition that keeps the accurate nearest-neighbor prior approximation and still yields a sparse covariance factor suitable for filtering. The prior precision is approximated by the forward-KL-optimal sparse inverse Cholesky factor on the nearest-neighbor pattern (Stage~1), the resulting posterior precision is factorized by incomplete Cholesky (Stage~2), and a sparse factor of the posterior \emph{covariance}, whose columns are multi-resolution basis functions with local supports, is recovered by a sparse approximate inverse \citep{Grote1997} that decouples into small, independent least-squares problems with closed-form solutions (Stage~3). This shift from conditional (autoregressive) to generative (moving-average) sparsity gives samples and forecast covariances without any global linear solve, at quasilinear cost and in an embarrassingly parallel way. Beyond filtering, it gives posterior variances and conditional simulations for any latent Gaussian model with conditionally independent observations. We show that the Stage-3 objective controls the reverse KL divergence, so the method can be viewed as a closed-form variational approximation in the spirit of \citet{Cao2023}, but over sparse covariance rather than sparse precision factors. We also show that HV is the special case of our pipeline in which Stages~2 and~3 are exact. An exact decomposition of the approximation error then explains why our method can nonetheless be considerably more accurate than HV at the same sparsity of the covariance factor: it trades a small, computable error in Stages~2 and~3 for a much smaller Stage-1 error. At the same sparsity, its conditioning sets are also much smaller than those of HV, so that it is at least as fast and, for large $n$, considerably faster. Because nearest-neighbor conditioning sets do not depend on a partition of the domain, our method also avoids the boundary artifacts of HV (Figure~\ref{fig:boundary}). Non-Gaussian data are handled through a Laplace approximation \citep{Jurek2020}.

Section~\ref{sec:methodology} introduces the three stages, their properties, the relation to HV, and the filter. Section~\ref{sec:numerical} compares our method with HV and low-rank approximations in simulations, Section~\ref{sec:application} applies it to satellite data, and Section~\ref{sec:conclusions} concludes. Proofs, further details, and additional results are given in Sections~\ref{app:proofs}--\ref{app:application} of the separate Supplementary Material, whose sections, equations, and figures are numbered with the prefix S.

\section{Methodology \label{sec:methodology}}

\subsection{Model and Overview}\label{sec:model}

We consider the spatio-temporal state-space model
\begin{equation}\label{eq:ssm}
\by_t\mid\bx_t\sim\textstyle\prod_{i=1}^{n} g_{t,i}(y_{t,i}\mid x_{t,i}),\qquad
\bx_t=\bH_t\bx_{t-1}+\bw_t,\quad \bw_t\sim\normal(\bfzero,\bQ_t),\qquad t=2,\dots,T,
\end{equation}
with $\bx_1\sim\normal(\bfmu_{1|0},\bfSigma_{1|0})$. The latent state $\bx_t=(x_{t,1},\dots,x_{t,n})^\top$ is a Gaussian field at locations $\bs_1,\dots,\bs_n$ in a domain $\domain\subset\mathbb R^d$, the evolution matrix $\bH_t$ is sparse, and the densities $g_{t,i}$ are log-concave, such as the Gaussian, Bernoulli--logistic, Poisson, or gamma distributions. Only $n_t\le n$ of the locations may be observed at time $t$.

\paragraph{Filtering recursion}
The filter alternates between a forecast and an update step. Suppose that the filtering distribution at time $t-1$ is $\bx_{t-1}\mid\by_{1:t-1}\sim\normal(\bfnu_{t-1},\bL_{t-1}\bL_{t-1}^\top)$. The forecast step then gives $\bx_t\mid\by_{1:t-1}\sim\normal(\bfmu,\bfSigma)$ with
\begin{equation}\label{eq:forecast}
\bfmu=\bH_t\bfnu_{t-1},\qquad \bfSigma=\bH_t\bL_{t-1}\bL_{t-1}^\top\bH_t^\top+\bQ_t ,
\end{equation}
and $(\bfmu,\bfSigma)=(\bfmu_{1|0},\bfSigma_{1|0})$ at $t=1$. For linear Gaussian models, this recursion is exactly the Kalman filter \citep{Kalman1960}; for non-Gaussian data and with the approximations introduced below, the filtering distributions are Gaussian approximations. The update step combines this forecast distribution, which acts as the prior, with the data $\by_t$. Every update is therefore an instance of the same problem, and in the remainder of this section we drop the time index: the prior is $\bx\sim\normal_n(\bfmu,\bfSigma)$, and the data are $y_i\mid\bx\stackrel{\mathrm{ind}}{\sim}g_i(y_i\mid x_i)$.

\paragraph{Gaussian data}
For Gaussian data, $y_i\mid\bx\sim\normal(x_i,d_i)$, the posterior is $\bx\mid\by\sim\normal(\bfnu,\bW^{-1})$ with
\begin{equation}\label{eq:posterior}
\bW=\bfSigma^{-1}+\bD^{-1},\qquad \bfnu=\bW^{-1}\big(\bfSigma^{-1}\bfmu+\bD^{-1}\by\big),
\end{equation}
where $\bD=\operatorname{diag}(d_1,\dots,d_n)$, and unobserved locations have $d_i=\infty$, i.e., a zero entry in $\bD^{-1}$.

\paragraph{Non-Gaussian data via Laplace approximation}
For non-Gaussian data, we follow \citet{Zilber2019} and \citet{Jurek2020} and use a Laplace approximation, computed by Newton--Raphson iterations that can be written as iterated pseudo-data updates. At the current iterate $\bx^{(k)}$, let $u_i=\partial_{x_i}\log g_i(y_i\mid x_i)$ and $d_i=-(\partial^2_{x_i}\log g_i(y_i\mid x_i))^{-1}$, both evaluated at $x_i^{(k)}$ (with $u_i=0$ and $d_i=\infty$ at unobserved locations), and define the pseudo-data $\bt^{(k)}=\bx^{(k)}+\bD^{(k)}\bu$ and pseudo-variances $\bD^{(k)}=\operatorname{diag}(d_1,\dots,d_n)$. The Newton step is then exactly the Gaussian posterior mean \eqref{eq:posterior} with $\by$ and $\bD$ replaced by $\bt^{(k)}$ and $\bD^{(k)}$:
\begin{equation}\label{eq:newton}
\bx^{(k+1)}=\big(\bfSigma^{-1}+(\bD^{(k)})^{-1}\big)^{-1}\big(\bfSigma^{-1}\bfmu+(\bD^{(k)})^{-1}\bt^{(k)}\big).
\end{equation}
At convergence, $\bx\mid\by\approx\normal(\bfnu,\bW^{-1})$, where $\bfnu$ is the posterior mode and $\bW=\bfSigma^{-1}+\bD^{-1}$ is evaluated at the final pseudo-variances. For Gaussian data, $\bt^{(k)}=\by$, $\bD^{(k)}=\bD$, and a single step gives the exact posterior. In both cases, the update returns a Gaussian distribution $\normal(\bfnu,\bW^{-1})$, and the precision $\bW$ differs from the prior precision only by a diagonal matrix.

\paragraph{Three-stage approximation}
To close the recursion, each update has to deliver the posterior mean $\bfnu$, which gives the next forecast mean $\bH\bfnu$, and a factor of the posterior \emph{covariance} $\bW^{-1}$, which gives the next forecast covariance $\bH\bW^{-1}\bH^\top+\bQ$ in \eqref{eq:forecast}. Exact computation of either costs $\order(n^3)$. The posterior mean requires only Stage~1 below, whereas the covariance factor requires all three stages, each of which computes a sparse triangular factor:
\begin{enumerate}
\item \emph{Stage~1} (Section~\ref{sec:stage1}): the prior precision is approximated by a sparse inverse Cholesky (Vecchia) factor, $\bfSigma^{-1}\approx\hU\hU^\top$, on a nearest-neighbor pattern. This yields the sparse approximate posterior precision $\tW=\hU\hU^\top+\bD^{-1}$, from which $\bfnu$ is computed by conjugate gradient.
\item \emph{Stage~2} (Section~\ref{sec:stage2}): the posterior precision is factorized by incomplete Cholesky, $\tW\approx\bV\bV^\top$, on the same pattern.
\item \emph{Stage~3} (Section~\ref{sec:stage3}): a sparse factor of the posterior covariance, $\hL\approx\bV^{-\top}$, is computed on a support pattern by a sparse approximate inverse.
\end{enumerate}
The result $\normal(\bfnu,\hL\hL^\top)$ has exactly the form assumed at the start of the recursion. Section~\ref{sec:sparsity} shows why the covariance factor $\hL$ makes the forecast step cheap, Section~\ref{sec:hv} analyzes the error of the three stages and the relation to HV, and Section~\ref{sec:filter} summarizes the resulting filter.

\subsection{Ordering and Sparsity Patterns}\label{sec:patterns}

\begin{figure}[t!]
    \centering
    \includegraphics[width=\textwidth]{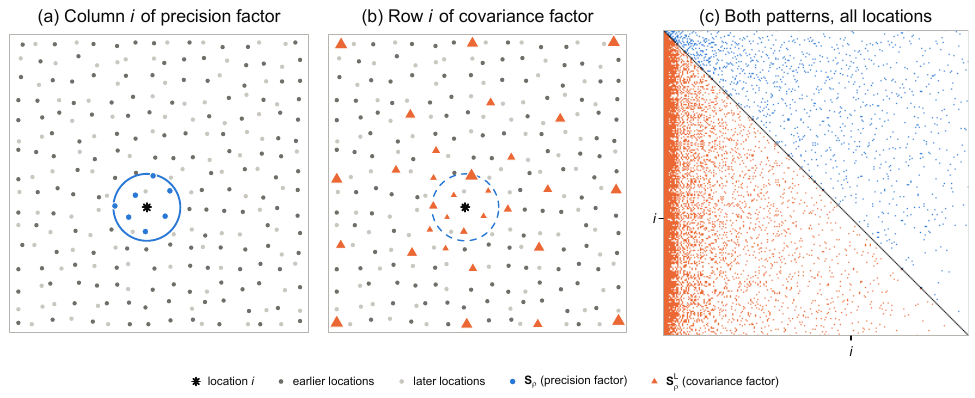}
    \caption{The two sparsity patterns for $n=256$ irregularly spaced locations (a randomly perturbed $16\times16$ grid) in maximin ordering, with $\rho=2$. Dark and light gray dots are the locations ordered before and after the fine-scale location $\bs_i$ (star), respectively. (a)~Column $i$ of $\Srho$, which contains the earlier locations within distance $\rho\ell_i$ of $\bs_i$ (circle), i.e., the conditioning set of $x_i$. (b)~Row $i$ of $\SA$, which contains the earlier locations $\bs_j$ with $\|\bs_i-\bs_j\|<\rho\ell_j$; the size of the triangles increases with $\ell_j$, and the dashed circle is the circle of (a). (c)~The patterns $\Srho$ (upper triangle) and $\SA$ (lower triangle) for all locations, with rows and columns in maximin order; the ticks mark column $i$ and row $i$.}
    \label{fig:patterns}
\end{figure}

Assume throughout that the locations are ordered according to a \emph{maximin ordering} \citep{Guinness2016a,Schafer2017}, in which each location maximizes the minimum distance to the previously ordered locations, so that the ordering fills the domain in a multi-resolution fashion. For each $i$, let $\ell_i=\min_{j<i}\|\bs_i-\bs_j\|$ be the distance to the nearest previously ordered location, with $\ell_1=\infty$. For a radius parameter $\rho\ge1$, we define the upper-triangular nearest-neighbor pattern and the lower-triangular support pattern
\begin{equation}\label{eq:conditioning}
\Srho=\{(j,i): j\le i,\ \|\bs_i-\bs_j\|<\rho\,\ell_i\},\qquad
\SA=\{(i,j): j\le i,\ \|\bs_i-\bs_j\|<\rho\,\ell_j\}.
\end{equation}
We write $\mathcal M(\bS)$ for the set of triangular matrices with sparsity pattern $\bS$, and denote the index sets of column $i$ of $\Srho$ and of row $i$ of $\SA$ by
\[
\cK_i=\{j:(j,i)\in\Srho\},\qquad \cJ_i=\{j:(i,j)\in\SA\},
\]
both of which contain $i$ as their largest element. Stages~1 and~2 use $\Srho$, and Stage~3 uses $\SA$. Column $i$ of $\Srho$ consists of the earlier locations within distance $\rho\ell_i$ of $\bs_i$, so that the radius is set by the later, finer-scale location. Column $j$ of $\SA$ consists of the later locations within distance $\rho\ell_j$ of $\bs_j$, so that the radius is set by the earlier, coarser-scale location. This column is the support of the $j$th basis function in Section~\ref{sec:sparsity}, hence the name support pattern. Because $\ell_j\ge\ell_i$ for $j\le i$, $\SA$ contains the transpose of $\Srho$. Row $i$ of $\SA$ consists of the earlier locations $\bs_j$ with $\|\bs_i-\bs_j\|<\rho\ell_j$. Since this radius grows with the scale of $\bs_j$, the row contains nearby fine-scale locations and increasingly distant coarse-scale locations, i.e., earlier locations at all scales. Figure~\ref{fig:patterns} compares column $i$ of $\Srho$ with row $i$ of $\SA$ for a fine-scale location $\bs_i$. If the locations are quasi-uniform in a domain of dimension $d$ (e.g., a regular grid), then $m_S=\max_i|\cK_i|=\order(\rho^d)$ and $m_L=\max_i|\cJ_i|=\order(\rho^d\log n)$ \citep{Schafer2017}. Each row of $\SA$ thus has only $\order(\rho^d\log n)$ entries, but the columns of $\SA$ associated with early-ordered (coarse-scale) locations may have up to $\order(n)$ entries.

\subsection{Stage 1: Prior Approximation via Forward KL}\label{sec:stage1}

Following \citet{Schafer2020}, the Vecchia approximation of the prior is the sparse inverse Cholesky factor that minimizes the forward Kullback--Leibler (KL) divergence,
\begin{equation}
\hU=\argmin_{\bU\in\mathcal M(\Srho)}\KL\big(\normal(\bfzero,\bfSigma)\,\|\,\normal(\bfzero,(\bU\bU^\top)^{-1})\big).
\end{equation}
The optimization decouples over the columns of $\bU$, and column $i$ has the closed-form solution
\begin{equation}\label{eq:vecchia}
\hU_{\cK_i,i}=\frac{[\bfSigma]_{\cK_i,\cK_i}^{-1}\,\mathbf e}{\big(\mathbf e^\top[\bfSigma]_{\cK_i,\cK_i}^{-1}\mathbf e\big)^{1/2}},
\end{equation}
where $\mathbf e=(0,\dots,0,1)^\top\in\mathbb R^{|\cK_i|}$ indicates the position of $i$, the largest element of $\cK_i$. Each column requires only the $|\cK_i|\times|\cK_i|$ submatrix of the prior covariance on the conditioning set $\cK_i$, so $\hU$ costs $\order(nm_S^3)$ and never requires the full matrix $\bfSigma$. By the screening effect \citep{Stein2002}, $\hU\hU^\top$ approximates $\bfSigma^{-1}$ accurately for a wide range of covariances, including Mat\'ern \citep[e.g.,][]{Guinness2016a,Katzfuss2017a,Schafer2020}. Theoretically, if the entries of $\bfSigma$ are given by the Green's function of an elliptic differential operator of order $2s>d$ with zero boundary conditions (which includes Mat\'ern covariances with $\nu+d/2\in\mathbb N$, up to boundary effects) at quasi-uniform locations, then $\KL\big(\normal(\bfzero,\bfSigma)\,\|\,\normal(\bfzero,(\hU\hU^\top)^{-1})\big)\le\epsilon$ for $\rho\ge C\log(n/\epsilon)$ \citep[][Sec.~3.3]{Schafer2020}. Under related conditions and fixed-domain asymptotics, \citet{Kang2024} show that conditioning sets of size $\order(\log^dn)$ suffice for this KL divergence to vanish as $n\to\infty$.

Replacing $\bfSigma^{-1}$ by $\hU\hU^\top$ in \eqref{eq:posterior} gives the sparse approximate posterior precision
\begin{equation}
\tW=\hU\hU^\top+\bD^{-1}.
\end{equation}
The posterior mean or mode is computed by the Newton iteration \eqref{eq:newton} with $\bfSigma^{-1}$ replaced by $\hU\hU^\top$ (one step for Gaussian data). To avoid overshooting, e.g., for Poisson data with large intensities, each step is halved until it does not decrease the log posterior $\sum_i\log g_i(y_i\mid x_i)-\tfrac12(\bx-\bfmu)^\top\hU\hU^\top(\bx-\bfmu)$, and we stop when $\max_i|x^{(k+1)}_i-x^{(k)}_i|<10^{-6}$. Each Newton step solves a sparse linear system by conjugate gradient \citep{Kundig2024}. $\hU$ is computed once per time point before the Newton iterations, and Stages~2 and~3 once after them, at the final pseudo-variances.

\subsection{Stage 2: Precision Factorization via Incomplete Cholesky}\label{sec:stage2}

The second stage computes an upper-triangular factor $\bV$ of the approximate posterior precision, $\tW\approx\bV\bV^\top$. To avoid fill-in, we use the incomplete Cholesky factorization $\bV=\operatorname{ichol}(\tW,\Srho)$ (IC0), which runs the Cholesky recursion only on $\Srho$. Using the larger pattern of $\hU\hU^\top$, which is contained in $\bS_{2\rho}$, can further increase the accuracy \citep[][Sec.~4.1]{Schafer2020}. Each of the $\order(nm_S)$ entries requires a sparse inner product of length at most $m_S$, so Stage~2 costs $\order(nm_S^2)$, and $\bV$ inherits the sparsity of $\hU$. For latent Vecchia models, \citet[][Secs.~4.1 and~5.2]{Schafer2020} found IC0 on $\Srho$ accurate over a wide range of $\rho$ and noise levels, and on the pattern of $\hU\hU^\top$ as accurate as exact Cholesky; no proof is available. However, $\bV$ is a factor of the posterior \emph{precision}, whereas the forecast step requires a factor of the posterior \emph{covariance}.

\subsection{Stage 3: Covariance Recovery via Sparse Approximate Inverse}\label{sec:stage3}

Since $\tW^{-1}\approx(\bV\bV^\top)^{-1}=\bV^{-\top}\bV^{-1}$, a lower-triangular approximation $\bL\approx\bV^{-\top}$ gives a covariance factor with $\bL\bL^\top\approx\tW^{-1}$. We compute it by the sparse approximate inverse (SPAI) method of \citet{Grote1997} with the Frobenius-norm objective, restricted to the support pattern:
\begin{equation}\label{eq:stage3}
\hL=\argmin_{\bL\in\mathcal M(\SA)}\|\bI-\bL\bV^\top\|_F^2 .
\end{equation}
The pattern $\SA$ is motivated by the structure of the target. The matrix $\bV^{-\top}$ is the lower-triangular Cholesky factor of the Stage-2 covariance $(\bV\bV^\top)^{-1}$, and for any covariance matrix $\bC=\bL\bL^\top$ with lower-triangular Cholesky factor $\bL$ and $\bx\sim\normal(\bfzero,\bC)$, we have $L_{ij}=\operatorname{Cov}(x_i,x_j\mid\bx_{1:j-1})/\operatorname{Var}(x_j\mid\bx_{1:j-1})^{1/2}$ for $j\le i$. Hence, $L_{ij}=0$ if and only if $x_i$ and $x_j$ are conditionally independent given $\bx_{1:j-1}$ \citep[][Sec.~3.2]{Schafer2017}. Under the maximin ordering, the earlier locations $\bs_1,\dots,\bs_{j-1}$ cover the domain with spacing $\ell_j$, so that, by the screening effect, $x_i$ and $x_j$ are nearly independent given $\bx_{1:j-1}$ unless $\|\bs_i-\bs_j\|$ is a small multiple of $\ell_j$, that is, unless $(i,j)\in\SA$. $\SA$ is the lower-triangular part of the pattern of \citet{Schafer2017}, who prove that, under the elliptic conditions of Section~\ref{sec:stage1}, the truncated Cholesky factor has Frobenius error at most $\epsilon$ for $\rho\ge C\log(n/\epsilon)$ (their Thm.~3.1); the diagonal $\bD^{-1}$ acts like a zeroth-order term and preserves this decay (their Sec.~4.3). Unlike incomplete Cholesky of the covariance matrix \citep{Schafer2017}, which needs posterior covariance entries and can break down, Stage~3 works with $\bV$ and always returns a valid covariance factor (Proposition~\ref{prop:klbound}).

\paragraph{Row-parallel closed-form solution}
The squared Frobenius norm is the sum of the squared norms of the rows, and row $i$ of $\bI-\bL\bV^\top$ is $(\mathbf e_i-\bV\boldsymbol\ell_i)^\top$, where $\boldsymbol\ell_i=\bL_{i,:}^\top$ is supported on $\cJ_i$. Hence \eqref{eq:stage3} decouples into $n$ independent least-squares problems $\min\|\mathbf e_i-\bV\boldsymbol\ell_i\|_2^2$, one per row. Let $\widehat\bV_i=\bV_{:,\cJ_i}$ be the columns of $\bV$ indexed by $\cJ_i$, let $\bG_i=\widehat\bV_i^\top\widehat\bV_i$ be their Gram matrix, and let $p_i=|\cJ_i|$ be the position of $i$ in $\cJ_i$. Because $\bV$ is upper triangular and $j\le i$ for all $j\in\cJ_i$, the right-hand side $\widehat\bV_i^\top\mathbf e_i$ of the normal equations of this least-squares problem has the single nonzero entry $V_{ii}$, at position $p_i$, so the nonzero entries of row $i$ are
\begin{equation}\label{eq:spai}
\hL_{i,\cJ_i}^\top=V_{ii}\,\bG_i^{-1}\mathbf e_{p_i}.
\end{equation}
Each entry of $\bG_i$ is an inner product of two columns of $\bV$ with at most $m_S$ nonzeros, so forming $\bG_i$ costs $\order(m_L^2m_S)$. With the Cholesky factorization $\bG_i=\bC_i\bC_i^\top$, which costs $\order(m_L^3)$, we have $\bG_i^{-1}\mathbf e_{p_i}=\bC_i^{-\top}\mathbf e_{p_i}/(\bC_i)_{p_ip_i}$, because $\bC_i^{-1}$ is lower triangular and $p_i$ is the last position. A single triangular solve thus gives row $i$, whose diagonal entry is $\widehat L_{ii}=V_{ii}/(\bC_i)_{p_ip_i}^2$. The total cost is $\order(nm_L^3)$, which is $\order\big(n(\rho^d\log n)^3\big)$ and hence quasilinear in $n$ for quasi-uniform locations, and all rows can be computed in parallel. Moreover, Stage~3 inverts the factor $\bV$ rather than $\tW$ itself, whose condition number is the square of that of $\bV$.

\paragraph{Variational perspective}
Stage~3 can be viewed as a variational approximation within the family of Gaussian distributions with sparse covariance factors,
\begin{equation}
\mathcal Q_{\mathrm{cov}}=\big\{\normal(\bfnu,\bB\bB^\top):\bB\in\mathcal M(\SA)\big\}.
\end{equation}
The natural objective is the reverse KL divergence to the target. For two Gaussians with a common mean,
\begin{equation}\label{eq:kl}
\KL\big(\normal(\bfnu,\bfSigma_Q)\,\|\,\normal(\bfnu,\bfSigma_P)\big)
=\tfrac12\Big[\operatorname{tr}(\bfSigma_P^{-1}\bfSigma_Q)-n-\log\det(\bfSigma_P^{-1}\bfSigma_Q)\Big].
\end{equation}
Minimizing the reverse KL divergence directly over $\mathcal Q_{\mathrm{cov}}$ also has a closed-form solution, but it decouples over the columns of $\bB$, and the columns associated with coarse-scale locations require solves with large submatrices of $\tW$, up to $\tW$ itself for the first column (Section~\ref{app:reversekl}). Proposition~\ref{prop:klbound} shows that the row-decoupled Frobenius objective \eqref{eq:stage3} instead controls the reverse KL divergence to the Stage-2 target, and that the quality of the Stage-3 solution can be assessed at no extra cost.

\begin{proposition}[Stage~3 controls the reverse KL]\label{prop:klbound}
Let $\bV\in\mathcal M(\Srho)$ with $V_{ii}>0$, let $P_{\bV}=\normal(\bfnu,(\bV\bV^\top)^{-1})$ be the Stage-2 target, and let $\kappa=\kappa_2(\bV)=\kappa_2(\bV\bV^\top)^{1/2}$, where $\kappa_2(\bA)=\|\bA\|_2\|\bA^{-1}\|_2=\sigma_{\max}(\bA)/\sigma_{\min}(\bA)$ is the spectral condition number, with $\sigma_{\max}(\bA)$ and $\sigma_{\min}(\bA)$ the largest and smallest singular values of $\bA$.
\begin{enumerate}
\item[(a)] The Stage-3 problem \eqref{eq:stage3} has the unique solution \eqref{eq:spai}. Its diagonal entries $\widehat L_{ii}=V_{ii}[\bG_i^{-1}]_{p_ip_i}$ are positive, so $\hL$ is nonsingular and $\normal(\bfnu,\hL\hL^\top)\in\mathcal Q_{\mathrm{cov}}$. Its residual is
\[
\varepsilon^2=\|\bI-\hL\bV^\top\|_F^2=\textstyle\sum_{i=1}^n\big(1-V_{ii}\widehat L_{ii}\big),\qquad 1-V_{ii}\widehat L_{ii}\in[0,1).
\]
\item[(b)] For any nonsingular lower-triangular $\bL$, with $Q_{\bL}=\normal(\bfnu,\bL\bL^\top)$ and $\varepsilon=\|\bI-\bL\bV^\top\|_F$,
\[
\KL(Q_{\bL}\|P_{\bV})=\tfrac12\textstyle\sum_{k=1}^n\big(\sigma_k^2-1-\log\sigma_k^2\big),
\]
where $\sigma_k$ are the singular values of $\bV^\top\bL$. If $\kappa\varepsilon\le\delta<1$, then
\[
\KL(Q_{\bL}\|P_{\bV})\;\le\;\tfrac12\big(1+(1-\delta)^{-2}\big)\,\kappa^2\varepsilon^2 .
\]
\end{enumerate}
\end{proposition}

\noindent
By (a), Stage~3 always returns a valid covariance factor, and the quality of each row can be read off from $V_{ii}\widehat L_{ii}$, with values close to one meaning near exactness. By (b), $\KL(Q_{\bL}\|P_{\bV})=0$ if and only if $\bV^\top\bL$ is orthogonal, and the bound is \emph{quadratic} in $\varepsilon$, simplifying to $\KL\lesssim\kappa^2\varepsilon^2$ for small residuals. Together, (a) and (b) give an a posteriori bound that requires only the diagonals of $\bV$ and $\hL$ and an estimate of $\kappa_2(\bV)$, e.g., from a few Lanczos iterations with sparse products and triangular solves with $\bV$, each costing $\order(nm_S)$. The residual is amplified by $\kappa_2(\bV)=\kappa_2(\tW)^{1/2}$ rather than by $\kappa_2(\tW)$, which makes precise why inverting $\bV$ is preferable to working with $\tW$. Finally, by Weyl's inequality, the eigenvalues of $\tW=\hU\hU^\top+\bD^{-1}$ lie in $[\lambda_{\min}(\hU\hU^\top)+\min_iD_{ii}^{-1},\,\lambda_{\max}(\hU\hU^\top)+\max_iD_{ii}^{-1}]$, so $\kappa$ is controlled by the conditioning of the prior precision and by the range of the noise precisions.

\paragraph{Relation to sparse precision families}
\citet{Cao2023} minimize the reverse KL divergence over the family $\mathcal Q_{\mathrm{prec}}=\{\normal(\bfnu,(\bB\bB^\top)^{-1}):\bB\in\mathcal M(\Srho)\}$ of sparse \emph{precision} factors. By \eqref{eq:kl}, this objective does not separate over $\bB$ and is optimized by stochastic gradient descent, whereas Stage~3 solves independent least-squares problems in closed form \eqref{eq:spai}. The price is a denser factor, with up to $m_L$ entries per row of $\SA$ instead of $m_S$ per column of $\Srho$.

\subsection{Conditional versus Generative Sparsity}\label{sec:sparsity}

The factors $\hU$ and $\bV$ of Stages~1 and~2 are sparse \emph{precision} factors, which correspond to an autoregressive (AR) representation in which each $x_i$ is regressed on a few previously ordered neighbors. The factor $\hL$ of Stage~3 is a sparse \emph{covariance} factor, which corresponds to a moving-average (MA), or generative, representation,
\begin{equation}
\bx=\bfnu+\hL\bz=\bfnu+\textstyle\sum_{j=1}^n z_j\hL_{:,j},\qquad \bz\sim\normal(\bfzero,\bI_n),
\end{equation}
in which the field is a weighted sum of the columns of $\hL$ with independent standard normal weights. The columns of $\hL$ thus act as spatial basis functions, and column $j$ is nonzero only at locations $\bs_i$, $i\ge j$, within distance $\rho\ell_j$ of $\bs_j$. Both types of factors encode conditional independence, but given different variables. By Proposition~\ref{prop:armap}(a) below, a zero entry $V_{ji}$, $j<i$, means that $x_i$ and $x_j$ are conditionally independent given all other previously ordered variables, whereas a zero entry $\widehat L_{ij}$ means that they are conditionally independent given only the variables ordered before $x_j$ (Section~\ref{sec:stage3}).

\begin{figure}[t!]
    \centering
    \includegraphics[width=\textwidth]{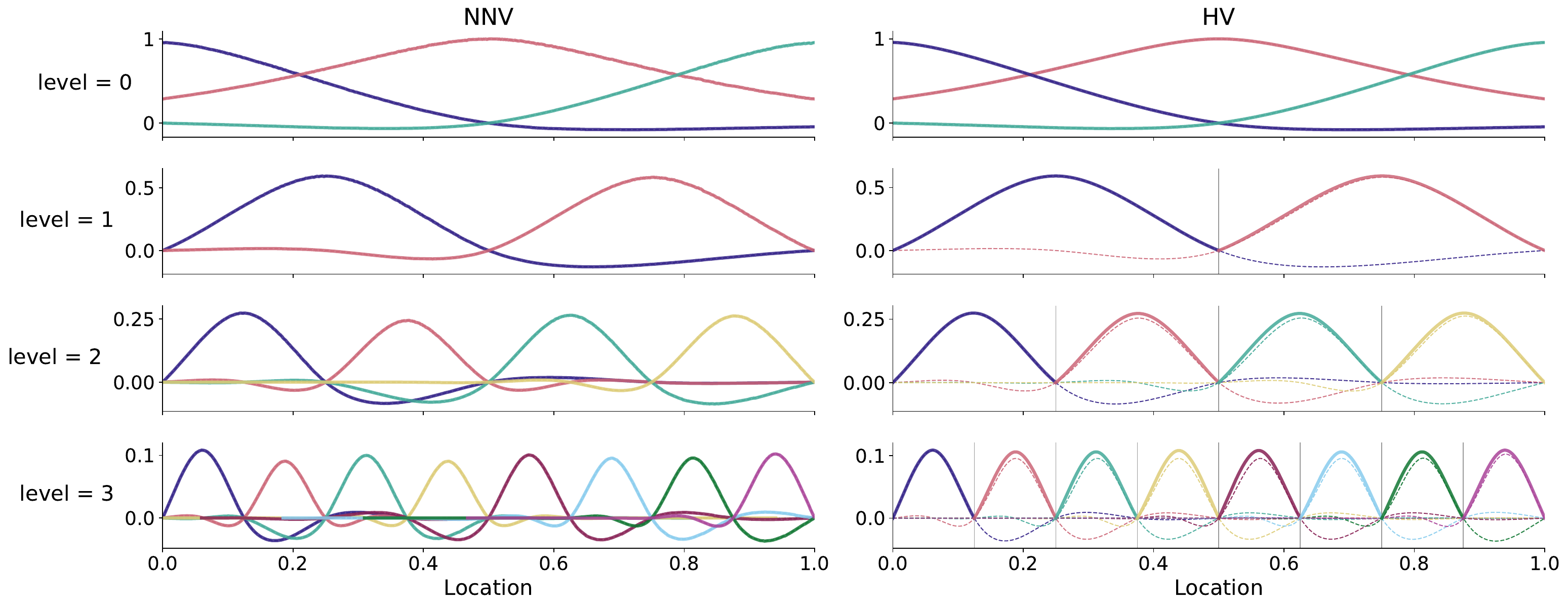}
    \caption{Implied prior basis functions of \NNV{} (left, columns of $\bL_{\text{prior}}$) and HV (right, columns of $\bL_{\mathrm{HV}}$) at the four coarsest resolution levels (rows, from coarse to fine), for $n=1{,}025$ points on $[0,1]$, a Mat\'ern covariance (smoothness $1.5$, range $0.2$) and a maximum conditioning-set size (Section~\ref{sec:setup}) of $16$ for both methods. Each HV column is shown only on its region, outside of which it is zero. Gray vertical lines mark the boundaries of the HV regions at each level. Colors distinguish the basis functions within a level. Thin dashed lines show the corresponding columns of the exact Cholesky factor of the prior covariance in the same ordering. In the left panels, they are covered by the \NNV{} basis functions.}
    \label{fig:basis_functions}
\end{figure}

Figure~\ref{fig:basis_functions} shows such basis functions for a Gaussian process on $[0,1]$. It compares the columns of the prior covariance factor $\bL_{\text{prior}}$, obtained by applying Stage~3 to $\hU$ (i.e., with $\bV=\hU$), with those of HV, $\bL_{\mathrm{HV}}=\hU_{\mathrm{HV}}^{-\top}$ (Section~\ref{sec:hv}); details are given in Section~\ref{app:basis}. Both have a multi-resolution structure: the basis functions at level~$0$ cover the whole domain, and their support and peak height shrink at finer levels. The nearest-neighbor basis functions are centered at their own locations, have similar shapes across the domain, and overlap. The HV basis functions are instead cut off at the boundaries of their regions, where they have kinks, so that those of neighboring regions at the same level do not overlap. The \NNV{} basis functions are visually indistinguishable from the exact ones (dashed lines), whereas HV omits the parts of the exact basis functions outside their regions. This difference translates into accuracy. The KL divergence from the exact prior to the Stage-1 approximation is $3.3\times10^{-2}$ for the nearest-neighbor pattern and $80.5$ for the hierarchical pattern. The cut-off basis functions of HV also cause artifacts along the partition boundaries (Section~\ref{sec:boundary}).

\begin{proposition}[Conditional (AR) versus generative (MA) sparsity]\label{prop:armap}
\mbox{}
\begin{enumerate}
\item[(a)] Let $\bV\in\mathcal M(\Srho)$ with $V_{ii}>0$, and $\bx\sim\normal(\bfnu,(\bV\bV^\top)^{-1})$. Then $\bz:=\bV^\top(\bx-\bfnu)\sim\normal(\bfzero,\bI_n)$ and
\[
p(x_i\mid\bx_{1:i-1})=\normal\big(\nu_i-V_{ii}^{-1}\textstyle\sum_{j\in\cK_i\setminus\{i\}}V_{ji}(x_j-\nu_j),\,V_{ii}^{-2}\big),
\]
so $p(\bx)=\prod_i p(x_i\mid\bx_{\cK_i\setminus\{i\}})$ and the log-density costs $\order(\operatorname{nnz}(\bV))$. A sample requires the sparse triangular solve $\bV^\top(\bx-\bfnu)=\bz$, which is inherently sequential. The same holds for the prior factor $\hU$ with $\bfnu=\bfmu$.
\item[(b)] Let $\bL\in\mathcal M(\SA)$ with $L_{ii}>0$ and $\bx=\bfnu+\bL\bz$, $\bz\sim\normal(\bfzero,\bI_n)$. Then $x_i=\nu_i+\sum_{j\in\cJ_i}L_{ij}z_j$, so a sample costs $\order(\operatorname{nnz}(\bL))=\order(n\,m_L)$ with all rows computed in parallel, and any covariance entry $\operatorname{Cov}(x_i,x_k)=\sum_{j\in\cJ_i\cap\cJ_k}L_{ij}L_{kj}$ costs $\order(m_L)$.
\item[(c)] In (b), $p(x_i\mid\bx_{1:i-1})=\normal\big(\nu_i+\sum_{j\in\cJ_i\setminus\{i\}}L_{ij}z_j,\;L_{ii}^2\big)$ with $\bz_{1:i-1}=\bL_{1:i-1,1:i-1}^{-1}(\bx-\bfnu)_{1:i-1}$. The conditional is sparse in $\bz$, but its dependence on $\bx$ runs through the rows of $\bL^{-1}$, which need not be sparse, so the density of $\bx$ need not admit a sparse factorization.
\end{enumerate}
\end{proposition}

\noindent
Part~(a) is the autoregressive view of a precision factor. The whitened vector $\bz=\bV^\top(\bx-\bfnu)$ consists of independent innovations, and row $i$ of this relation expresses $x_i$ as a linear regression on its $|\cK_i|-1$ nearest previously ordered neighbors plus an independent innovation with variance $V_{ii}^{-2}$. The density therefore factorizes into sparse conditionals, and likelihood evaluation requires one sparse matrix--vector product. Generating a sample, however, requires inverting this relation, $\bx=\bfnu+\bV^{-\top}\bz$, by forward substitution, in which $x_i$ can only be computed after all $x_j$ with $j\in\cK_i\setminus\{i\}$. Part~(b) is the moving-average view of a covariance factor. Each $x_i$ is an explicit weighted sum of at most $m_L$ innovations, so sampling and the evaluation of any covariance entry reduce to sparse inner products that can be computed for all $i$ simultaneously. Part~(c) shows what is lost in exchange: the conditional distribution of $x_i$ given the past is sparse in the innovations but not in the past values $\bx_{1:i-1}$, because recovering the innovations requires $\bL^{-1}$, which is generally dense. The two representations are thus dual. Sparse precision factors give cheap densities and expensive samples, while sparse covariance factors give cheap samples and covariances and expensive densities. Filtering requires covariances, as the following result shows, which is why Stage~3 targets the covariance factor.

\begin{proposition}[The forecast step requires no global linear solve]\label{prop:forecast}
Suppose $\bx_t\mid\by_{1:t}\sim\normal(\bfnu_t,\bL_t\bL_t^\top)$ with $\bL_t\in\mathcal M(\SA)$, and $\bx_{t+1}=\bH\bx_t+\bw_{t+1}$, where $\bw_{t+1}\sim\normal(\bfzero,\bQ)$ is independent of $\bx_t$ and $\by_{1:t}$, $\bH$ has at most $c_H$ nonzeros per row, and each entry of $\bQ$ can be evaluated in $\order(1)$ time (e.g., $\bQ$ is sparse or given by a covariance function). Then $\bx_{t+1}\mid\by_{1:t}\sim\normal(\bH\bfnu_t,\;\bF\bF^\top+\bQ)$ with $\bF=\bH\bL_t$, and:
\begin{enumerate}
\item[(a)] each row of $\bF$ has at most $c_Hm_L$ nonzeros, and $\bF$ is computed in $\order(n\,c_Hm_L)$ time;
\item[(b)] any entry of $\bfSigma_{t+1|t}=\bF\bF^\top+\bQ$ costs $\order(c_Hm_L)$ time;
\item[(c)] the Stage-1 factor $\hU_{t+1}$ for prior $\bfSigma_{t+1|t}$ and pattern $\Srho$ needs only the submatrices $[\bfSigma_{t+1|t}]_{\cK_i,\cK_i}$, $i=1,\dots,n$, whose distinct entries all lie in $\bS_{2\rho}\cup\bS_{2\rho}^\top$ and number $\order(n\rho^d)$, so that $\hU_{t+1}$ is computed in $\order(nm_S^3+n\,c_H\rho^dm_L)$ time without forming any dense $n\times n$ matrix;
\item[(d)] if, in addition, the locations are quasi-uniform and the evolution is local in the sense that $H_{kl}\neq0$ only if $\|\bs_k-\bs_l\|\le\rho\ell_n$, then each row of $\bF$ has $\order(\rho^d\log n)$ nonzeros irrespective of $c_H$, and the time in (c) is $\order(n\rho^{3d}+n\rho^{2d}\log n)$.
\end{enumerate}
\end{proposition}

\noindent
Proposition~\ref{prop:forecast} explains how the generative factor is used in the forecast step \eqref{eq:forecast}. In MA form, the forecast amounts to the substitution $\bx_{t+1}=\bH\bfnu_t+(\bH\bL_t)\bz+\bw_{t+1}$, so the forecast covariance factor $\bF=\bH\bL_t$ is obtained by a sparse matrix--matrix product. Part~(a) shows that $\bF$ stays sparse, because each of its rows combines at most $c_H$ rows of $\bL_t$. Part~(b) shows that any single entry of the (dense) forecast covariance can be evaluated on demand from two sparse rows of $\bF$. Part~(c) observes that the next Stage-1 approximation \eqref{eq:vecchia} needs only the small submatrices $[\bfSigma_{t+1|t}]_{\cK_i,\cK_i}$, which share most of their entries, so that only $\order(n\rho^d)$ distinct entries have to be computed; this closes the loop of the filter at quasilinear cost. Part~(d) shows that, for local evolution such as a finite-difference discretization of a partial differential equation, the rows of $\bL_t$ that are combined in a row of $\bF$ share most of their supports, so that $\bF$ is not denser than $\bL_t$ up to a constant and the cost does not grow with $c_H$. By contrast, with only the precision factor $\bV$ of Proposition~\ref{prop:armap}(a), each entry $[\bH\tW^{-1}\bH^\top]_{kl}=(\bV^{-1}\bh_k)^\top(\bV^{-1}\bh_l)$, where $\bh_k^\top$ is the $k$th row of $\bH$, requires triangular solves whose results are generally dense, so assembling the entries needed in Proposition~\ref{prop:forecast}(c) costs $\order(n\cdot\operatorname{nnz}(\bV))$, i.e., quadratic in $n$. This, rather than sampling, is the decisive advantage of the generative factor for filtering.

\paragraph{Connection to nonlinear generative models}
Just as sparse inverse Cholesky factors extend to nonlinear autoregressive models \citep[e.g.,][]{Katzfuss2021}, the sparse covariance factor is the linear special case of nonlinear generative models such as inverse autoregressive flows \citep{Kingma2016}, which replace the linear combinations $\bL_{i,:}\bz$ by nonlinear maps $f_i(\bz_{\cJ_i})$. For nonlinear models, however, analytical propagation through the evolution breaks down, and nonlinear extensions therefore lead naturally to ensemble-based filters rather than closed-form updates \citep[e.g.,][]{Johnson2026}.

\subsection{Error Decomposition and Hierarchical Vecchia as a Special Case}\label{sec:hv}

The three stages target the successive distributions $P_{\mathrm{post}}=\normal(\bfnu,\bW^{-1})$, $\widetilde P=\normal(\bfnu,\tW^{-1})$ and $P_{\bV}=\normal(\bfnu,(\bV\bV^\top)^{-1})$. For symmetric positive definite $\bA,\bB$ and a covariance matrix $\bfSigma_Q$, let
\[
\Delta(\bA,\bB;\bfSigma_Q)=\tfrac12\big[\operatorname{tr}\big((\bA-\bB)\bfSigma_Q\big)-\log\det(\bA\bB^{-1})\big].
\]
Then any $Q=\normal(\bfnu,\bfSigma_Q)$ satisfies the exact identity
\begin{equation}\label{eq:errdecomp}
\KL(Q\|P_{\mathrm{post}})
=\underbrace{\KL(Q\|P_{\bV})}_{\text{Stage 3}}
+\underbrace{\Delta(\tW,\bV\bV^\top;\bfSigma_Q)}_{\text{Stage 2}}
+\underbrace{\Delta(\bW,\tW;\bfSigma_Q)}_{\text{Stage 1}},
\end{equation}
which follows by telescoping (see the proof of Proposition~\ref{prop:hv}). Each term vanishes if the corresponding stage is exact. The Stage-3 term is nonnegative and bounded by Proposition~\ref{prop:klbound}. The Stage-2 and Stage-1 terms measure how much the KL divergence changes when the precision matrix of the target is replaced by its approximation, and they can in principle be negative, i.e., one approximation error may partially offset another. Two features of \eqref{eq:errdecomp} are important. First, $\bW-\tW=\bfSigma^{-1}-\hU\hU^\top$ does not depend on $\bD$, so the Stage-1 term depends on the data only through $\bfSigma_Q$ and otherwise measures the error of the \emph{prior} approximation alone. Second, $\Delta(\bW,\tW;\tW^{-1})=\KL(\widetilde P\|P_{\mathrm{post}})$, i.e., if $Q$ were exactly the Stage-1 posterior, the Stage-1 term would be the KL divergence between the Stage-1 posterior and the true posterior. The identity assumes a common mean $\bfnu$. If $Q$ has a different mean $\bfnu_Q$, the additional term $\tfrac12(\bfnu_Q-\bfnu)^\top\bW(\bfnu_Q-\bfnu)$ appears, which concerns the computation of the posterior mean in Stage~1 and not the covariance factorization.

\begin{remark}[Bounds on the Stage-1 and Stage-2 terms]\label{rem:bounds}
For given $\bD$, the Stage-1 term at the Stage-1 posterior is bounded by the error of the Stage-1 prior approximation,
\[
\Delta(\bW,\tW;\tW^{-1})=\KL(\widetilde P\|P_{\mathrm{post}})\le\KL\big(\normal(\bfmu,(\hU\hU^\top)^{-1})\,\|\,\normal(\bfmu,\bfSigma)\big),
\]
for any noise variances (Section~\ref{app:proofs}). For other $Q$, the Stage-1 term differs from the left-hand side by $\tfrac12\operatorname{tr}\big((\bW-\tW)(\bfSigma_Q-\tW^{-1})\big)$, the trace of the product of the Stage-1 error $\bW-\tW$ and the combined Stage-2 and Stage-3 error $\bfSigma_Q-\tW^{-1}$. If $\bfSigma$ satisfies the conditions of Section~\ref{sec:stage1} under which $\KL\big(\normal(\bfzero,\bfSigma)\,\|\,\normal(\bfzero,(\hU\hU^\top)^{-1})\big)\le\epsilon$ for $\rho\ge C\log(n/\epsilon)$ \citep{Schafer2020}, then for $\epsilon\le0.1$, the right-hand side above is at most $2\epsilon$ (Section~\ref{app:proofs}), so that the Stage-1 term decays exponentially in $\rho$. No comparable guarantee is available for the Stage-2 term, i.e., for the incomplete Cholesky factorization of $\tW$ (Section~\ref{sec:stage2}).
\end{remark}

\paragraph{Hierarchical Vecchia}
The HV filter \citep{Jurek2020} and the related multi-resolution approximation \citep{Katzfuss2015} use conditioning sets based on a recursive partitioning of the spatial domain. Under this block-hierarchical ordering, the conditioning sets of the prior coincide with the full ancestor sets, so the precision factors $\hU$ and $\bV$ and the covariance factor $\hL$ share the same block-sparse footprint. If we apply our three stages with the hierarchical pattern instead of the nearest-neighbor pattern, Stages~2 and~3 become exact.

\begin{proposition}[HV is the exact special case]\label{prop:hv}
Let $\mathcal C_i$ be the HV conditioning sets of \citet[eq.~(4)]{Jurek2020}, which are recalled in Section~\ref{app:hvsets}, and use the upper-triangular pattern $\bS_{\mathrm{HV}}=\{(j,i):\bs_j\in\mathcal C_i\cup\{\bs_i\}\}$ in Stages~1--2 and its transpose $\bS_{\mathrm{HV}}^\top$ in Stage~3. For symmetric positive definite $\bA$, let $\operatorname{rchol}(\bA)$ denote the unique upper-triangular matrix with positive diagonal such that $\bA=\operatorname{rchol}(\bA)\operatorname{rchol}(\bA)^\top$, let $\operatorname{chol}(\bA)$ denote the usual lower-triangular Cholesky factor, and let $\operatorname{ichol}(\bA,\bS)$ denote the incomplete version of $\operatorname{rchol}$ restricted to the pattern $\bS$. Then Stage~1 returns the HV prior factor $\hU_{\mathrm{HV}}$, and
\begin{enumerate}
\item[(a)] Stage~2 is exact: $\bV=\operatorname{ichol}(\tW,\bS_{\mathrm{HV}})=\operatorname{rchol}(\tW)$;
\item[(b)] Stage~3 is exact: $\hL=\bV^{-\top}=\operatorname{chol}(\tW^{-1})$;
\item[(c)] consequently, the HV approximation $Q_{\mathrm{HV}}=\normal(\bfnu,\hL\hL^\top)$ satisfies $\KL(Q_{\mathrm{HV}}\|P_{\mathrm{post}})=\KL(\widetilde P_{\mathrm{HV}}\|P_{\mathrm{post}})$, where $\widetilde P_{\mathrm{HV}}=\normal(\bfnu,(\hU_{\mathrm{HV}}\hU_{\mathrm{HV}}^\top+\bD^{-1})^{-1})$; that is, the entire error of HV is the Stage-1 error of the hierarchical pattern.
\end{enumerate}
\end{proposition}

\noindent
The hierarchical pattern has a special closure property, established by \citet{Jurek2020}: the exact Cholesky factor of the posterior precision has no fill-in outside $\bS_{\mathrm{HV}}$, and the exact Cholesky factor of the posterior covariance has no fill-in outside $\bS_{\mathrm{HV}}^\top$. Because incomplete Cholesky and SPAI are exact whenever the exact answer lies in the prescribed pattern, Stages~2 and~3 reproduce the exact factors, and by \eqref{eq:errdecomp} the error of HV reduces to the Stage-1 term at the Stage-1 posterior. Exactness of Stages~2 and~3 does \emph{not} make HV the more accurate method. HV places its entire error budget in Stage~1, by restricting the conditioning sets to a block partitioning that guarantees exact inversions, whereas the nearest-neighbor sparse Cholesky decomposition accepts small, controlled and parallelizable errors in Stages~2 and~3, bounded a posteriori by Proposition~\ref{prop:klbound}, in order to use the much more accurate nearest-neighbor pattern in Stage~1. By \eqref{eq:errdecomp} and Proposition~\ref{prop:hv}, the triple approximation $Q_{\mathrm{NN}}$ is more accurate than HV if and only if
\[
\KL(Q_{\mathrm{NN}}\|P_{\bV})+\Delta(\tW_{\mathrm{NN}},\bV\bV^\top;\bfSigma_{Q_{\mathrm{NN}}})
+\Delta(\bW,\tW_{\mathrm{NN}};\bfSigma_{Q_{\mathrm{NN}}})
\;<\;\KL(\widetilde P_{\mathrm{HV}}\|P_{\mathrm{post}}).
\]
The first term on the left is controlled by Proposition~\ref{prop:klbound}, and the second is small in practice, because incomplete Cholesky is accurate for latent Vecchia models (Section~\ref{sec:stage2}). Because $\bW-\tW=\bfSigma^{-1}-\hU\hU^\top$, the comparison of the third term with the right-hand side is essentially a comparison of prior approximations (Remark~\ref{rem:bounds}), for which nearest-neighbor conditioning sets are much more accurate than hierarchical ones of the same size \citep[][Figs.~3 and~6]{Katzfuss2017a}. The numerical experiments in Section~\ref{sec:numerical} confirm that the inequality holds in practice. At the same number of nonzeros per row of the covariance factor, the conditioning sets of HV are also several times larger than those of \NNV{} (Section~\ref{sec:setup}), so that HV is not only less accurate but also, for large $n$, slower.

\subsection{The Filtering Algorithm}\label{sec:filter}

Combining the forecast step of Proposition~\ref{prop:forecast} with the three stages gives the filter for the state-space model \eqref{eq:ssm}, summarized in Algorithm~\ref{alg:filter}.

\begin{algorithm}[htbp]
\caption{Nearest-neighbor sparse Cholesky filter (\NNV)}\label{alg:filter}
\KwInput{maximin-ordered locations, patterns $\Srho$ and $\SA$; $\bfmu_{1|0}$, $\bfSigma_{1|0}$; $\bH_t$, $\bQ_t$ and data $\by_t$, $t=1,\dots,T$}
\For{$t=1,\dots,T$}{
\emph{Forecast:} if $t>1$, set $\bfmu=\bH_t\bfnu_{t-1}$ and $\bF_t=\bH_t\hL_{t-1}$, and compute only the entries $[\bF_t\bF_t^\top+\bQ_t]_{\cK_i,\cK_i}$, $i=1,\dots,n$, of the forecast covariance $\bfSigma$; if $t=1$, use $\bfmu_{1|0}$ and $\bfSigma_{1|0}$\;
\emph{Stage 1:} compute $\hU_t$ by \eqref{eq:vecchia} from these entries\;
\emph{Update:} compute the posterior mean or mode $\bfnu_t$ and the final pseudo-variances $\bD_t$ by the iterations \eqref{eq:newton} with $\bfSigma^{-1}$ replaced by $\hU_t\hU_t^\top$, solving each step by conjugate gradient\;
\emph{Stage 2:} compute $\bV_t=\operatorname{ichol}(\tW_t,\Srho)$ with $\tW_t=\hU_t\hU_t^\top+\bD_t^{-1}$\;
\emph{Stage 3:} compute $\hL_t$ row by row by \eqref{eq:spai}\;
}
\KwOut{filtering distributions $\bx_t\mid\by_{1:t}\approx\normal(\bfnu_t,\hL_t\hL_t^\top)$, $t=1,\dots,T$}
\end{algorithm}

For quasi-uniform locations and fixed $\rho$, the forecast step and Stages~1--3 have a time complexity that is quasilinear in $n$, dominated by the $\order(nm_L^3)=\order(n\log^3n)$ cost of Stage~3. Each conjugate-gradient iteration costs $\order(n)$, and no dense $n\times n$ matrix is ever formed. For nonlinear evolution $\bx_t=\mathcal E(\bx_{t-1})+\bw_t$, we proceed as \citet{Jurek2020}, following the extended Kalman filter \citep{Jazwinski1970}. We set $\bfmu=\mathcal E(\bfnu_{t-1})$ and replace $\bH_t$ by the Jacobian of $\mathcal E$ at $\bfnu_{t-1}$.

\paragraph{Parameter inference}
The filter also provides an approximation of the likelihood of unknown parameters $\bftheta$ in $\bfmu_{1|0}$, $\bfSigma_{1|0}$, $\bH_t$, $\bQ_t$, or $g_{t,i}$, at negligible additional cost. The likelihood factorizes as $p(\by_{1:T}\mid\bftheta)=\prod_{t=1}^Tp(\by_t\mid\by_{1:t-1},\bftheta)$, and each factor is the normalizing constant of the update at time $t$. Writing $\log p(\by_t\mid\by_{1:t-1})=\log p(\by_t\mid\bx_t)+\log p(\bx_t\mid\by_{1:t-1})-\log p(\bx_t\mid\by_{1:t})$ at $\bx_t=\bfnu_t$, and replacing the forecast distribution by its Stage-1 approximation $\normal(\bfmu,(\hU_t\hU_t^\top)^{-1})$ and the filtering distribution by $\normal(\bfnu_t,(\bV_t\bV_t^\top)^{-1})$, we obtain
\[
\log p(\by_t\mid\by_{1:t-1},\bftheta)\approx\textstyle\sum_{i}\log g_{t,i}(y_{t,i}\mid\nu_{t,i})+\sum_{i=1}^n\log\widehat U_{t,ii}-\sum_{i=1}^n\log V_{t,ii}-\tfrac12\big\|\hU_t^\top(\bfnu_t-\bfmu)\big\|^2,
\]
where the first sum is over the observed locations. For Gaussian data, this is the exact log-likelihood under the Stage-1 forecast distribution if Stage~2 is exact; for non-Gaussian data, it is a Vecchia--Laplace approximation of the integrated likelihood \citep{Zilber2019}. It requires only the diagonals of $\hU_t$ and $\bV_t$ and one sparse matrix--vector product. The approximate likelihood can be maximized numerically over static parameters, with one run of the filter per evaluation, or it can provide the particle weights in a Rao-Blackwellized particle filter for static or time-varying parameters, as described for the multi-resolution and HV filters by \citet[][Sec.~6]{Jurek2018} and \citet[][Sec.~S2]{Jurek2020}.


\section{Numerical Experiments \label{sec:numerical}}

The $n$ least-squares problems of Stage~3 are independent, so all rows can be computed in parallel. Our implementation of Algorithm~\ref{alg:filter} is written in Python, compiled with Numba, and runs on the $32$ threads of an Intel Core i9-14900 CPU; it never forms a dense $n\times n$ matrix. The same code also runs HV and LR (Proposition~\ref{prop:hv}), with exact Stages~2 and~3 within the hierarchical pattern, so that all comparisons in Sections~\ref{sec:sim_spatial}, \ref{sec:sim_linear}, \ref{sec:time} and~\ref{sec:application} use the same implementation on the same hardware.

\subsection{Setup}\label{sec:setup}

Our simulations follow \citet[][Sec.~6]{Jurek2020}. The latent field has zero mean and exponential covariance $\exp(-\|\bs_i-\bs_j\|/0.15)$ on a regular $40\times40$ grid ($n=1{,}600$) on $\domain=[0,1]^2$, and it is observed through one of four data models: Gaussian with noise variance $\tau^2=0.2$, Bernoulli--logistic, Poisson with log link, and gamma with shape $2$ and log link. We compare \NNV{} (Algorithm~\ref{alg:filter}) with HV and its low-rank special case LR \citep{Jurek2020}, all in the CPU implementation of Section~\ref{sec:filter}, and with the exact (Laplace) posterior and Kalman filter, computed with dense matrices. For all three methods, $\cs$ denotes the median conditioning-set size, i.e., the median over $i$ of the number of locations on which $x_i$ is conditioned, not counting $\bs_i$ itself. For \NNV{}, this number is $|\cK_i|-1$, because $\cK_i$ in Section~\ref{sec:patterns} contains $i$, and $\cs$ is determined by $\rho$. For HV, it is $|\mathcal C_i|$, where the conditioning set $\mathcal C_i$ of Section~\ref{app:hvsets} does not contain $\bs_i$, and for LR, $\cs$ is the number of knots, on which every other location is conditioned. To separate the errors of the three stages (Section~\ref{sec:hv}), we also report \NNV{}~+ IC0, i.e., Stages~1--2 with exact $\bV^{-\top}$, and the Stage-1 approximation alone, for which Stages~2 and~3 are exact.

The methods are matched on the average number of nonzeros per row of the posterior covariance factor, $\sum_{i=1}^n|\cJ_i|/n$, which the filter stores and propagates and which determines the cost of the forecast step (Proposition~\ref{prop:forecast}). We use five levels with about $40$ to $120$ nonzeros per row; the settings are given in Section~\ref{app:numdetails}. Following \citet{Jurek2020}, we report the difference in log score \citep[dLS;][]{Gneiting2014}, $\mathrm{dLS}=\log p_{\mathrm{exact}}(\bx^{\mathrm{true}}\mid\by)-\log q(\bx^{\mathrm{true}}\mid\by)$, of the approximate posterior or filtering density $q$ of the entire field, whose expectation for Gaussian data is the KL divergence from the exact to the approximate distribution, and the relative root mean squared prediction error of the posterior mean, $\mathrm{RRMSPE}=\mathrm{RMSPE}(\text{method})/\mathrm{RMSPE}(\text{exact})$, both averaged over $100$ simulated data sets. Lower is better, with $\mathrm{dLS}=0$ and $\mathrm{RRMSPE}=1$ for the exact method.

\subsection{Spatial-Only Comparison}\label{sec:sim_spatial}

\begin{figure}[t!]
    \centering
    \includegraphics[width=\textwidth]{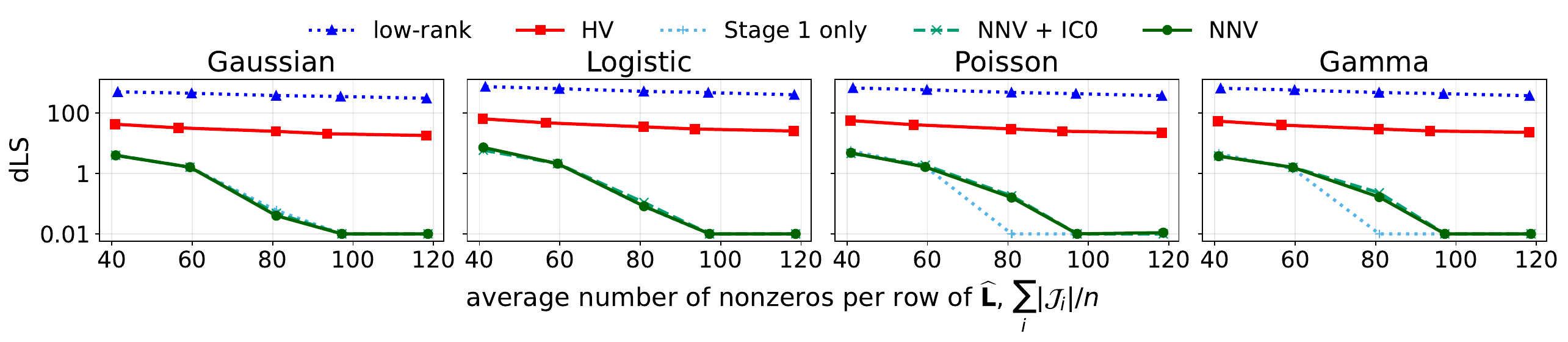}
    \caption{Spatial-only comparison ($n=1{,}600$; Section~\ref{sec:sim_spatial}): dLS against the average number of nonzeros per row of the posterior covariance factor, $\sum_i|\cJ_i|/n$ (Section~\ref{sec:setup}), for \NNV{}, \NNV{}~+ IC0 (exact Stage~3), the Stage-1 approximation alone (exact Stages~2 and~3), HV and low-rank. Columns correspond to the Gaussian, logistic, Poisson and gamma data models. Scores are averaged over $100$ simulated data sets. The dLS axis is on a log scale, and dLS values below $0.01$ (Monte Carlo error, including slightly negative values) are shown as $0.01$. The corresponding RRMSPE is shown in Figure~\ref{fig:rmspe_spatial}.}
    \label{fig:spatial}
\end{figure}

At every sparsity level and for all four data models, \NNV{} is considerably more accurate than HV, and both are far more accurate than LR (Figure~\ref{fig:spatial}). The dLS of \NNV{} is essentially zero from about $100$ nonzeros per row on, whereas that of HV decreases only slowly. The dLS of HV is about $9$ to $25$ times that of \NNV{} at the two sparsest levels and more than $170$ times at the three denser ones, and that of LR is more than $100$ times that of \NNV{} at every level. The curves of \NNV{}, \NNV{}~+ IC0 and the Stage-1 approximation coincide, so that Stages~2 and~3 add negligible error, and the advantage of \NNV{} comes from the much more accurate Stage-1 approximation, as anticipated by \eqref{eq:errdecomp}. In addition, the posterior means of HV show artifacts along its partition lines (Figure~\ref{fig:boundary} for logistic data, and Figure~\ref{fig:boundary_poisson} for Poisson data; Section~\ref{sec:boundary}). The RMSPE of the posterior means leads to the same ordering (Figures~\ref{fig:rmspe_spatial}--\ref{fig:rmspe_large_n}): at $n=1{,}600$, the RMSPE of \NNV{} is within about $0.4\%$ of that of the exact method in the spatial-only and the filtering comparison, whereas that of HV is up to about $3.5\%$ and that of LR up to about $27\%$ higher, and for large $n$, the RMSPE of HV is about $2$ to $3\%$ and that of LR about $20$ to $110\%$ above that of \NNV{}.

\subsection{Filtering with Advection--Diffusion Dynamics}\label{sec:sim_linear}

\begin{figure}[t!]
    \centering
    \includegraphics[width=\textwidth]{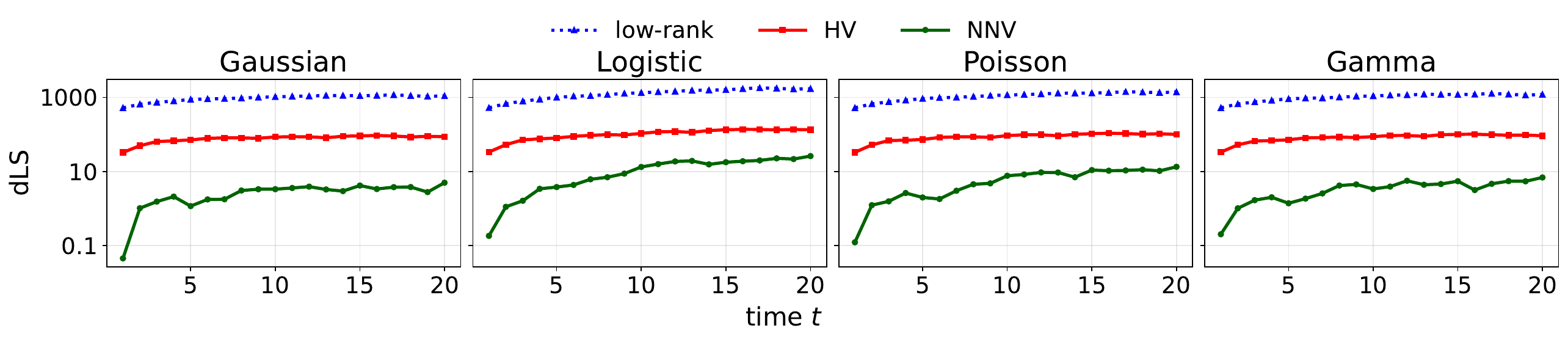}
    \caption{Linear advection--diffusion filtering ($n=1{,}600$; Section~\ref{sec:sim_linear}): dLS of the filtering distributions as a function of time $t=1,\dots,20$ for \NNV{}, HV and low-rank, averaged over $100$ simulated data sets, at the sparsity level with about $80$ nonzeros per row of the covariance factor (matched across methods to within $5\%$; Section~\ref{app:numdetails}). Columns correspond to the Gaussian, logistic, Poisson and gamma data models. The dLS axis is on a log scale. The corresponding RRMSPE is shown in Figure~\ref{fig:rmspe_linear}.}
    \label{fig:linear}
\end{figure}

Following \citet[][Sec.~6.3]{Jurek2020}, the latent field evolves according to
\[
\frac{\partial x(\bs,t)}{\partial t}=\alpha\Big(\frac{\partial^2 x}{\partial s_x^2}+\frac{\partial^2 x}{\partial s_y^2}\Big)
+\beta\Big(\frac{\partial x}{\partial s_x}+\frac{\partial x}{\partial s_y}\Big)+w(\bs,t),
\]
with $\alpha=4\times10^{-5}$ and $\beta=10^{-2}$, discretized by centered finite differences on the same grid, which gives an evolution matrix $\bH$ with at most five nonzeros per row. The model errors $w$ and the initial state have the exponential covariance above, $\bfSigma_{1|0}=\bQ_t$. At each of $T=20$ time points, a random $10\%$ of the locations are observed, with noise variance $\tau^2=0.25$ for Gaussian data. At every time point and for all data models, \NNV{} is considerably more accurate than HV, and LR has by far the largest errors (Figure~\ref{fig:linear}). The errors of all methods grow over the first time points, as the error of each update is propagated by $\bH$, but \NNV{} accumulates much less error than HV and LR. Averaged over the $20$ time points, the dLS of HV is about $8$ to $28$ times and that of LR about $100$ to $350$ times that of \NNV{}, depending on the data model. The same ordering holds at all sparsity levels (Figure~\ref{fig:linear_sparsity}).

\subsection{Computation Time}\label{sec:time}

\begin{figure}[t!]
    \centering
    \includegraphics[width=\textwidth]{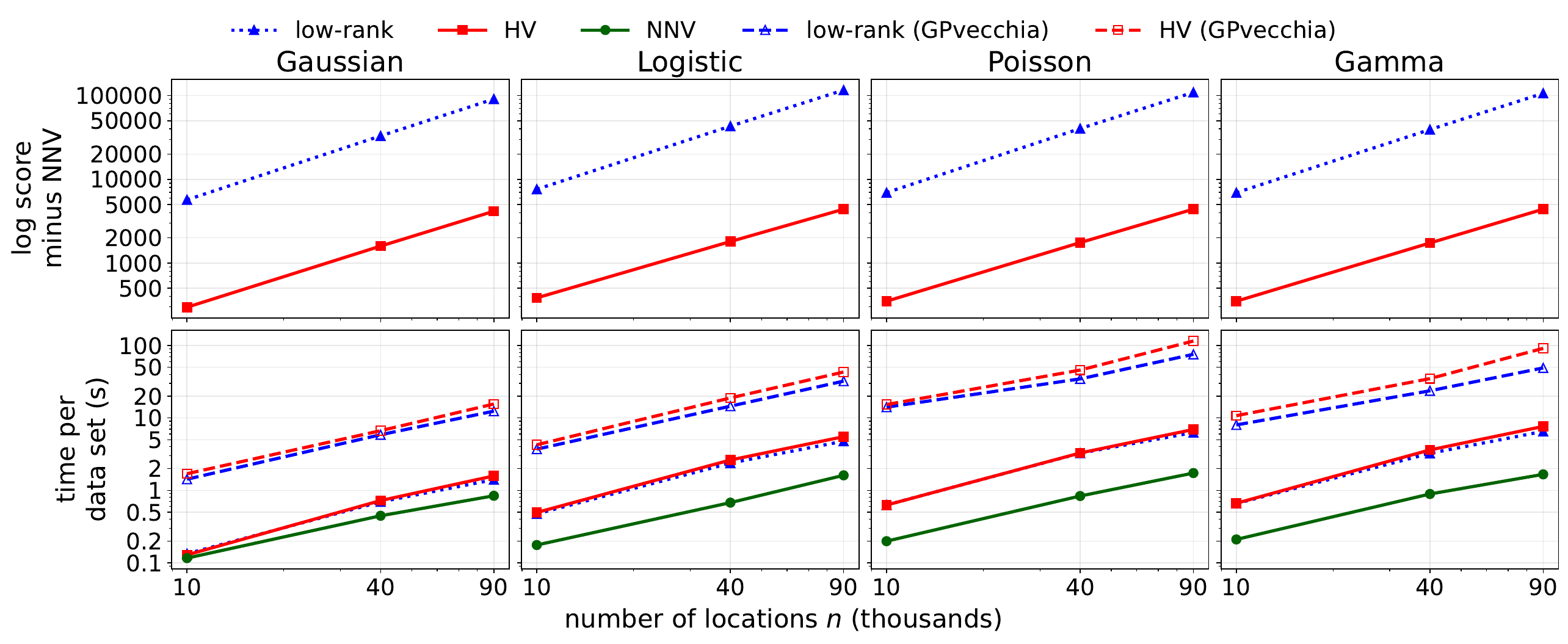}
    \caption{Spatial-only comparison for large $n$ (Section~\ref{sec:time}), with all methods at about $80$ nonzeros per row of the posterior covariance factor: log score of HV and low-rank minus that of \NNV{} (top row) and computation time per data set of all three methods (bottom row), against the number of locations $n$. Dashed lines with open markers show the times of HV and low-rank in GPvecchia, which exclude Stage~3. Columns correspond to the Gaussian, logistic, Poisson and gamma data models. Values are averaged over $10$ simulated data sets, and all axes are on a log scale. The RMSPE of HV and low-rank relative to \NNV{} is shown in Figure~\ref{fig:rmspe_large_n}.}
    \label{fig:time_spatial}
\end{figure}

We repeat the spatial-only comparison on grids with $n=10{,}000$, $40{,}000$ and $90{,}000$ locations, with all methods at about $80$ nonzeros per row and $10$ data sets per grid and data model; the settings are given in Section~\ref{app:numdetails}. The times include Stages~1--3 and all Newton iterations, but not the ordering and the sparsity patterns. At the same sparsity, \NNV{} is the most accurate method for every $n$ and data model and at least as fast as HV and LR (Figure~\ref{fig:time_spatial}), and its time grows approximately linearly in $n$. For Gaussian data at $n=10{,}000$, all three methods are about equally fast, but the advantage of \NNV{} grows with $n$; at $n=90{,}000$, \NNV{} is about twice as fast as HV and LR for Gaussian data and three to five times as fast for non-Gaussian data. Since sparser HV and LR approximations are faster but less accurate (Figure~\ref{fig:spatial}), an HV or LR approximation that runs as fast as \NNV{} would be even less accurate, so that \NNV{} is also the most accurate method at the same computation time. HV is slower mainly because, at the same sparsity of $\hL$, it needs about ten times larger conditioning sets than \NNV{}. HV and LR in the R package GPvecchia give the same log scores but are considerably slower than in our implementation (dashed lines). The filtering experiment of \citet[][Sec.~6.4]{Jurek2020} with $n=90{,}000$ leads to the same conclusions (Figure~\ref{fig:time_filter}).

\FloatBarrier


\section{Application to Satellite Data \label{sec:application}}

We apply the filter to total precipitable water vapor (TPW) measured by the Microwave Integrated Retrieval System (MIRS), as in Section~7 of \citet{Jurek2020} \citep[see also][]{Katzfuss2014}: the inner $247\times247$ cells ($n=61{,}009$) of the western $263\times263$ part of the grid and $T=11$ consecutive days, on which between $8\%$ and $60\%$ of the cells are observed. We hold out a random $10\%$ of the observations at each time point.

We model $y_{t,i}=\mu(\bs_i)+x_{t,i}+\epsilon_{t,i}$ with $\epsilon_{t,i}\sim\normal(0,\tau^2)$, where the coordinates $\bs_i$ are the grid indices scaled to $[0,1]^2$, and the mean $\mu$ is linear in the coordinates and fitted to the training data by least squares. The anomalies $\bx_t$ follow the advection--diffusion model of Section~\ref{sec:sim_linear}, and, as in the simulations, $\bfSigma_{1|0}$ and $\bQ_t$ are exponential covariance matrices with a common range $\lambda$ and variances $\sigma_0^2$ and $\sigma_Q^2$. The parameters are estimated by maximizing the approximate filtering likelihood of Section~\ref{sec:filter} on the full grid, using \NNV{} with $\rho=3.5$ and the training data. The estimates are $\tau^2=0.070$ for the noise variance, $\alpha=4.4\times10^{-8}$ for the diffusion coefficient, $\beta=1.6\times10^{-4}$ for the advection coefficient, $\lambda=0.2$ (about $49$ grid cells) for the range, and $\sigma_0^2=54.3$ and $\sigma_Q^2=10.2$ for the variances. All methods are matched at about $300$ nonzeros per row of $\hL_t$ (Section~\ref{sec:setup}), which corresponds to $\rho=5$ for \NNV{}, a requested conditioning-set size of $393$ for HV, and $\cs{}=307$ for LR. In both the parameter estimation and the filter, Stage~2 of \NNV{} runs incomplete Cholesky on the pattern of $\hU\hU^\top$ instead of $\Srho$ (Section~\ref{sec:stage2}). We use $\rho=5$ rather than a smaller value because, for smaller $\rho$, small Stage-3 errors accumulate in large regions without data and make the interval widths of \NNV{} visibly noisy. The predictive distribution of a held-out $y_{t,i}$ is $\normal\big(\mu(\bs_i)+\nu_{t,i},\,\|\hL_{t,i,:}\|^2+\tau^2\big)$. As in Section~\ref{sec:setup}, we report the dLS and RRMSPE relative to the exact filter, here of the held-out data, and we compare the filtering means of the approximations with those of the exact filter. We chose the size of this grid so that the exact Kalman filter, which stores the dense $n\times n$ filtering covariance matrix (about $30$\,GB), is just about feasible as a benchmark. Its memory grows quadratically in $n$ and its time per step up to cubically, so that it is infeasible on substantially larger grids, especially for parameter estimation, which requires many runs of the filter (Section~\ref{app:application}).

\begin{table}[t!]
    \centering\small
    \caption{TPW application: held-out RRMSPE and dLS relative to the exact Kalman filter (Section~\ref{sec:setup}), CRPS \citep{Gneiting2007}, coverage (PICP) and mean width (MPIW) of the $95\%$ prediction intervals, averaged over the $11$ time points, and total computation time in seconds for all $11$ filtering steps, for the exact filter, \NNV{}, HV and low-rank.}
    \label{tab:application}
    \begin{tabular}{lrrrrrr}
\toprule
method & RRMSPE & dLS & CRPS & PICP & MPIW & total time (s) \\
\midrule
exact & 1.000 & 0.000 & 0.283 & 0.965 & 2.79 & 2204 \\
\NNV{} & 1.001 & 0.000 & 0.283 & 0.965 & 2.79 & 366 \\
HV & 1.045 & 0.033 & 0.292 & 0.964 & 2.85 & 1247 \\
LR & 2.592 & 0.978 & 0.782 & 0.950 & 6.89 & 660 \\
\bottomrule
\end{tabular}

\end{table}

\begin{figure}[t!]
    \centering
    \includegraphics[width=\textwidth]{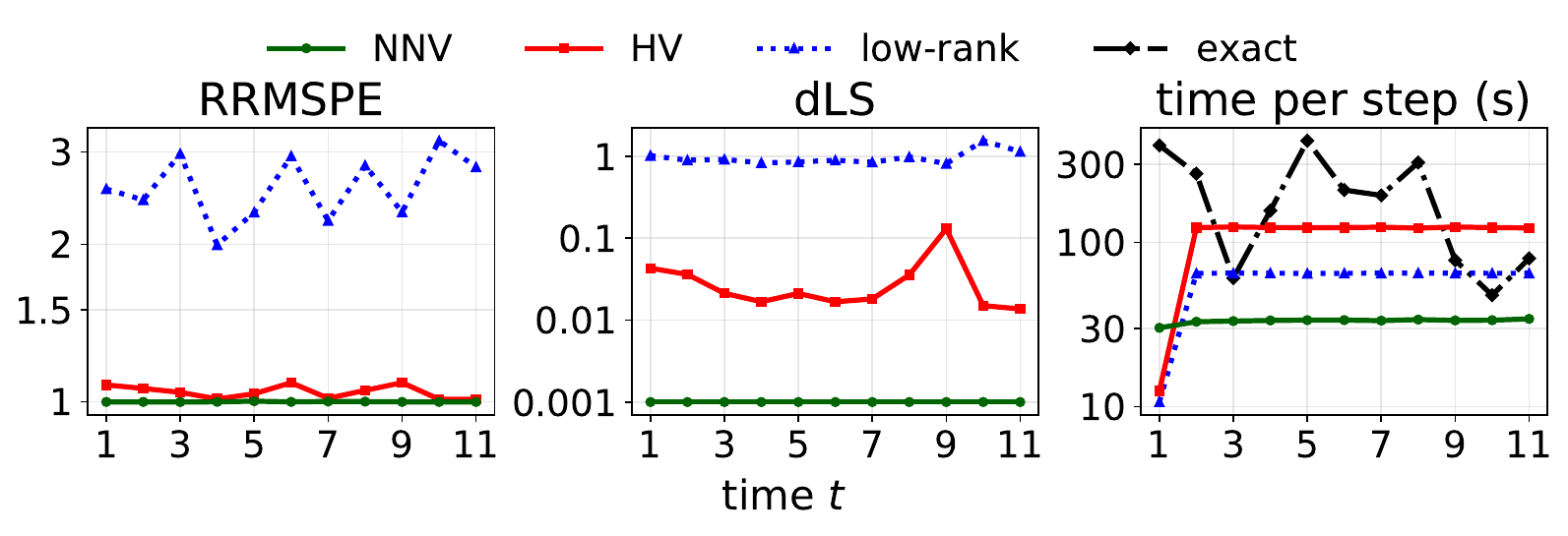}
    \caption{TPW application: held-out RRMSPE (left) and dLS (middle) of \NNV{}, HV and low-rank relative to the exact Kalman filter, and computation time per filtering step (right) of all four filters, over time, all on a log scale. dLS values below $0.001$ (including slightly negative values) are shown as $0.001$.}
    \label{fig:application_exact}
\end{figure}

\begin{figure}[t!]
    \centering
    \figureorplaceholder{width=\textwidth}{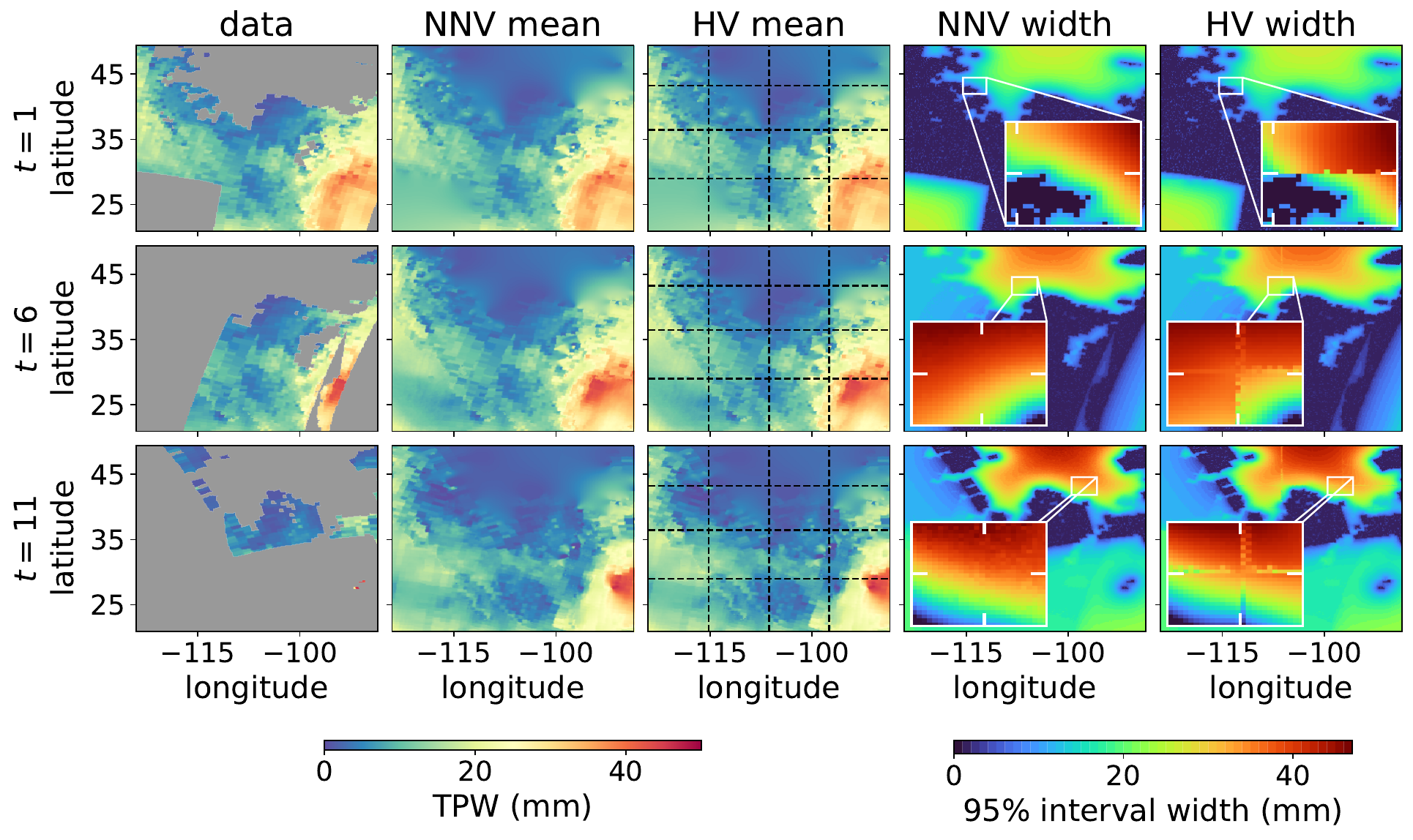}
    \caption{TPW application at $t=1$, $6$ and $11$ (rows): observations (left column, including the held-out ones; gray cells are unobserved), filtering means of \NNV{} and HV (common color scale), and widths of the pixelwise $95\%$ prediction intervals of \NNV{} and HV (common color scale with discrete $1$-mm bands). The filter does not use the randomly held-out $10\%$ of the observations, which therefore appear in the width maps as isolated cells with larger widths within the observed areas. Dashed lines in the HV mean maps show the coarsest partition boundaries of HV. The insets in the width maps magnify a region (white box) crossed by partition boundaries of HV (at $t=6$ and $11$, centered on a corner where two boundaries meet), marked by white ticks at the inset edges, on a stretched color scale common to the two insets of each row. The corresponding maps of low-rank are in Figure~\ref{fig:application_lowrank}.}
    \label{fig:application_maps}
\end{figure}

\NNV{} is essentially indistinguishable from the exact Kalman filter at the held-out data at every time point, whereas HV is noticeably and LR much less accurate (Table~\ref{tab:application} and Figure~\ref{fig:application_exact}). Compared directly with the filtering means and standard deviations of the exact filter, \NNV{} is also the closest at every time point {(Figure~\ref{fig:application_exact_rmsd})}, both at cells with data, where {its root-mean-square differences are about $8$ to $40$ times (means) and $65$ to $770$ times (standard deviations) smaller than those of HV}, and at cells without data, where the approximation errors are largest {and those of \NNV{} are about $1.6$ to $10$ times (means) and $2$ to $19$ times (standard deviations) smaller than those of HV}. \NNV{} is about three times as fast as HV and almost twice as fast as LR, and about six times as fast as the exact filter. The prediction intervals of \NNV{} and HV have close to nominal coverage. The filtering means of \NNV{} and HV are similar (Figure~\ref{fig:application_maps}), whereas LR loses small-scale detail and shows spurious features far from the data (Figure~\ref{fig:application_lowrank}). The interval widths of HV jump at its partition boundaries, whereas those of \NNV{} do not.



\section{Conclusions \label{sec:conclusions}}

We have proposed a nearest-neighbor sparse Cholesky decomposition for filtering in high-dimensional spatio-temporal state-space models. It approximates the prior precision by the forward-KL-optimal sparse inverse Cholesky factor on a nearest-neighbor pattern, factorizes the posterior precision by incomplete Cholesky, and recovers a sparse posterior covariance factor, whose columns are local multi-resolution basis functions, by a sparse approximate inverse. This shift from conditional to generative sparsity gives the forecast covariances without any global linear solve, so that the whole filter runs at quasilinear cost and in an embarrassingly parallel way. The last stage has closed-form solutions, and its objective controls the reverse KL divergence.

Hierarchical Vecchia is the special case of our pipeline in which Stages~2 and~3 are exact, at the price of a much larger Stage-1 error; our approach instead accepts small, computable errors in Stages~2 and~3 in exchange for a much more accurate prior approximation. In the numerical experiments, for Gaussian and non-Gaussian data, \NNV{} was close to the exact (Laplace) filter, at the same sparsity of the posterior covariance factor it was considerably more accurate than HV and low-rank filters and at least as fast as both (for $90{,}000$ locations, about twice as fast for Gaussian data and three to five times as fast for non-Gaussian data), and it showed no artifacts along partition boundaries. In the analysis of satellite measurements of total precipitable water vapor on $61{,}009$ grid cells, chosen so that the exact Kalman filter was just about feasible for comparison, \NNV{} was essentially indistinguishable from the exact filter and more accurate than HV and LR at randomly held-out observations, and it was about three times as fast as HV and almost twice as fast as LR.

Future work includes nonlinear evolution via the extended Kalman filter (Section~\ref{sec:filter}), other latent Gaussian models such as spatial generalized linear mixed models, and retrospective smoothing.

\if0\blind
{
\section*{Acknowledgments}

This work was partially supported by NASA AIST-23 grant 80NSSC26K0595 and by the Office of the Vice Chancellor for Research and Graduate Education at the University of Wisconsin--Madison with funding from the Wisconsin Alumni Research Foundation. We would like to thank Marcin Jurek and Jian Cao for helpful comments and discussions.
} \fi

\section*{Disclosure Statement}

The authors report there are no competing interests to declare.

\section*{Declaration of Generative AI Use}

During the preparation of this work, the authors used Claude (Anthropic) to assist with editing the manuscript and with writing and debugging code. The authors reviewed and edited all output and take full responsibility for the content of the article.

%

\bigskip
\begin{center}
{\large\bfseries SUPPLEMENTARY MATERIAL}
\end{center}

\noindent The Supplementary Material follows the references. It contains proofs, the reverse-KL optimal sparse covariance factor, the conditioning sets of HV, details of Figure~\ref{fig:basis_functions} and of the numerical experiments, additional numerical results (boundary effects of HV, prediction errors, and filtering at all sparsity levels and for large $n$), and additional results of the application, including details of the exact Kalman filter and the differences of the approximate filtering distributions from the exact ones.

\noindent Python and R code to reproduce all numerical results, tables, and figures of the article and the Supplementary Material is available at \url{https://github.com/katzfuss-group/NNV}, whose README also explains how to obtain the MIRS data of Section \ref{sec:application}.

\bibliographystyle{agsm}
\begingroup
\renewcommand{\bf}{\bfseries}
\renewcommand{\newline}{}
\def\harvardurl#1{}
\bibliography{mendeley,additionalrefs}
\endgroup

\end{document}


\maketitle

\noindent Sections, figures and equations in this supplement are numbered with the prefix S; all other references are to the main article.

\section{Proofs \label{app:proofs}}

Throughout, we use the following elementary facts. For matrices of compatible sizes,
\[
\|\bA\bB\bC\|_F\le\|\bA\|_2\|\bB\|_F\|\bC\|_2.
\]
 A triangular matrix with positive diagonal is nonsingular, its inverse is triangular of the same type, and each of its leading principal submatrices is nonsingular. If $\bT$ is lower triangular, then $(\bT\bv)_{1:i}=\bT_{1:i,1:i}\bv_{1:i}$ for every vector $\bv$ and every $i$.

\subsection*{Proof of Proposition~\ref{prop:klbound}}

We first show that the Stage-3 problem separates over the rows of $\bL$. The squared Frobenius norm of a matrix is the sum of the squared Euclidean norms of its rows, and row $i$ of $\bI-\bL\bV^\top$ is $(\mathbf e_i-\bV\boldsymbol\ell_i)^\top$ with $\boldsymbol\ell_i=\bL_{i,:}^\top$. Hence
\[
\|\bI-\bL\bV^\top\|_F^2=\textstyle\sum_{i=1}^n\|\mathbf e_i-\bV\boldsymbol\ell_i\|_2^2 .
\]
The constraint $\bL\in\mathcal M(\SA)$ only requires $\boldsymbol\ell_i$ to be supported on $\cJ_i$, separately for each $i$, so the sum is minimized by minimizing each summand separately.

Fix $i$ and let $\bu=(\boldsymbol\ell_i)_{\cJ_i}\in\mathbb R^{|\cJ_i|}$ denote the free entries of row $i$. Because $\boldsymbol\ell_i$ vanishes outside $\cJ_i$, we have $\bV\boldsymbol\ell_i=\widehat\bV_i\bu$, and the $i$th summand becomes the unconstrained least-squares objective
\[
f_i(\bu)=\|\mathbf e_i-\widehat\bV_i\bu\|_2^2=1-2\,\bu^\top\widehat\bV_i^\top\mathbf e_i+\bu^\top\bG_i\bu .
\]
Since $\bV$ is upper triangular with positive diagonal, it is nonsingular, so the columns of $\widehat\bV_i$ are linearly independent and $\bu^\top\bG_i\bu=\|\widehat\bV_i\bu\|_2^2>0$ for $\bu\ne\bfzero$, i.e., $\bG_i\succ0$. Hence $f_i$ is strictly convex, and its unique minimizer solves the normal equations $\bG_i\bu=\widehat\bV_i^\top\mathbf e_i$. The entries of $\widehat\bV_i^\top\mathbf e_i$ are $V_{ij}$, $j\in\cJ_i$, and since every $j\in\cJ_i$ satisfies $j\le i$ and $\bV$ is upper triangular, the only nonzero entry is $V_{ii}$, at position $p_i$. Therefore $\widehat\bu=V_{ii}\bG_i^{-1}\mathbf e_{p_i}$, which is \eqref{eq:spai}.

To prove (a), note that $\widehat L_{ii}$ is the $p_i$th entry of $\widehat\bu$, so $\widehat L_{ii}=V_{ii}[\bG_i^{-1}]_{p_ip_i}>0$, because the diagonal entries of $\bG_i^{-1}\succ0$ are positive. The lower-triangular matrix $\hL$ is thus nonsingular. Substituting $\widehat\bu$ into $f_i$ and using $\widehat\bu^\top\bG_i\widehat\bu=\widehat\bu^\top\widehat\bV_i^\top\mathbf e_i=V_{ii}^2[\bG_i^{-1}]_{p_ip_i}$, we obtain
\[
f_i(\widehat\bu)=1-V_{ii}^2[\bG_i^{-1}]_{p_ip_i}=1-V_{ii}\widehat L_{ii},
\]
which lies in $[0,1)$, since $f_i$ is a squared norm and $V_{ii}\widehat L_{ii}>0$. Summing over $i$ gives the expression for $\varepsilon^2$.

For (b), let $\bL$ be any nonsingular lower-triangular matrix and let $\bN=\bV^\top\bL$, with singular values $\sigma_1\ge\dots\ge\sigma_n>0$. By the cyclic property of the trace and the multiplicativity of the determinant, $\operatorname{tr}(\bV\bV^\top\bL\bL^\top)=\|\bN\|_F^2=\sum_k\sigma_k^2$ and $\det(\bV\bV^\top\bL\bL^\top)=\det(\bN)^2=\prod_k\sigma_k^2$. Substituting these into \eqref{eq:kl} with $\bfSigma_Q=\bL\bL^\top$ and $\bfSigma_P^{-1}=\bV\bV^\top$, we obtain the identity
\[
\KL(Q_{\bL}\|P_{\bV})=\tfrac12\Big[\textstyle\sum_k\sigma_k^2-n-\sum_k\log\sigma_k^2\Big]=\tfrac12\sum_kg(\sigma_k),\qquad g(\sigma)=\sigma^2-1-\log\sigma^2 .
\]
To bound the right-hand side, we relate $\bN-\bI$ to the Stage-3 residual $\bL\bV^\top-\bI$. The two matrices are similar, since $\bV^\top(\bL\bV^\top-\bI)\bV^{-\top}=\bV^\top\bL-\bI=\bN-\bI$, and hence
\[
\|\bN-\bI\|_2\le\|\bN-\bI\|_F\le\|\bV^\top\|_2\,\|\bL\bV^\top-\bI\|_F\,\|\bV^{-\top}\|_2=\kappa\varepsilon\le\delta,
\]
where $\kappa=\kappa_2(\bV)=\kappa_2(\bV\bV^\top)^{1/2}$, because the singular values of $\bV\bV^\top$ are the squares of those of $\bV$. By Weyl's inequality for singular values, $|\sigma_k-1|\le\|\bN-\bI\|_2\le\delta$, so that $\sigma_k\ge1-\delta>0$ for all $k$, and by Mirsky's theorem, $\sum_k(\sigma_k-1)^2\le\|\bN-\bI\|_F^2\le\kappa^2\varepsilon^2$. It remains to bound each $g(\sigma_k)$ by a multiple of $(\sigma_k-1)^2$. We have $g(1)=g'(1)=0$ and $g''(\sigma)=2+2\sigma^{-2}$, so by Taylor's theorem with Lagrange remainder, $g(\sigma)=\tfrac12g''(\xi)(\sigma-1)^2$ for some $\xi$ between $1$ and $\sigma$. If $\sigma\ge1-\delta$, then $\xi\ge1-\delta$, and therefore $g(\sigma)\le\big(1+(1-\delta)^{-2}\big)(\sigma-1)^2$. Combining these results,
\[
\KL(Q_{\bL}\|P_{\bV})\le\tfrac12\big(1+(1-\delta)^{-2}\big)\textstyle\sum_k(\sigma_k-1)^2\le\tfrac12\big(1+(1-\delta)^{-2}\big)\kappa^2\varepsilon^2 .\qquad\qed
\]

\subsection*{Proof of Proposition~\ref{prop:armap}}

We first prove (a). The vector $\bz=\bV^\top(\bx-\bfnu)$ is a linear transformation of a Gaussian vector with mean $\bfzero$ and covariance $\bV^\top(\bV\bV^\top)^{-1}\bV=\bI_n$, so its entries are independent $\normal(0,1)$ variables. Since $\bV^\top$ is lower triangular with positive diagonal, $\bz_{1:i-1}=(\bV^\top)_{1:i-1,1:i-1}(\bx-\bfnu)_{1:i-1}$ with an invertible leading block. Thus $\bx_{1:i-1}$ and $\bz_{1:i-1}$ determine each other, and $z_i$, being independent of $\bz_{1:i-1}$, is also independent of $\bx_{1:i-1}$. Row $i$ of $\bz=\bV^\top(\bx-\bfnu)$ reads $z_i=\sum_{j\le i}V_{ji}(x_j-\nu_j)$, and because column $i$ of $\bV$ is supported on $\cK_i$, solving for $x_i$ gives
\[
x_i=\nu_i-\frac{1}{V_{ii}}\sum_{j\in\cK_i\setminus\{i\}}V_{ji}(x_j-\nu_j)+\frac{z_i}{V_{ii}} .
\]
Conditionally on $\bx_{1:i-1}$, the sum is fixed and $z_i\sim\normal(0,1)$ is independent of it, which gives the stated conditional distribution. This conditional depends on $\bx_{1:i-1}$ only through $\bx_{\cK_i\setminus\{i\}}$, so it equals $p(x_i\mid\bx_{\cK_i\setminus\{i\}})$, and the chain rule $p(\bx)=\prod_ip(x_i\mid\bx_{1:i-1})$ yields the sparse factorization. For the costs, note that $\log p(\bx)=\sum_i\log V_{ii}-\tfrac12\|\bz\|_2^2-\tfrac n2\log(2\pi)$ requires a single sparse matrix--vector product, costing $\order(\operatorname{nnz}(\bV))$. Sampling, in contrast, requires solving $\bV^\top(\bx-\bfnu)=\bz$ by forward substitution, and by the autoregressive form above, $x_i$ can only be computed after all $x_j$ with $j\in\cK_i\setminus\{i\}$, which creates a chain of sequential dependencies through the ordering. The same arguments apply verbatim to $\hU$ and $\bfmu$.

For (b), since $L_{ij}=0$ for $j\notin\cJ_i$, row $i$ of $\bx=\bfnu+\bL\bz$ is $x_i=\nu_i+\sum_{j\in\cJ_i}L_{ij}z_j$. Each $x_i$ is thus an inner product of length $|\cJ_i|\le m_L$ that depends on $\bz$ only, so all rows can be computed in parallel, at total cost $\sum_i|\cJ_i|=\operatorname{nnz}(\bL)\le nm_L$. Moreover, $\operatorname{Cov}(\bx)=\bL\bL^\top$, so $\operatorname{Cov}(x_i,x_k)=\sum_jL_{ij}L_{kj}$, where only the indices $j\in\cJ_i\cap\cJ_k$ contribute. With sorted index lists, this sum is computed in $\order(|\cJ_i|+|\cJ_k|)=\order(m_L)$ operations.

Finally, for (c), $\bL$ is lower triangular with positive diagonal, so, as in (a), $(\bx-\bfnu)_{1:i-1}=\bL_{1:i-1,1:i-1}\bz_{1:i-1}$ with an invertible leading block, and $z_i$ is independent of $\bx_{1:i-1}$. Conditioning $x_i=\nu_i+\sum_{j\in\cJ_i\setminus\{i\}}L_{ij}z_j+L_{ii}z_i$ on $\bx_{1:i-1}$ fixes $\bz_{1:i-1}=\bL_{1:i-1,1:i-1}^{-1}(\bx-\bfnu)_{1:i-1}$ and leaves $L_{ii}z_i\sim\normal(0,L_{ii}^2)$, which gives the stated conditional. Written in terms of $\bx$, however, the conditional mean involves the rows of $\bL_{1:i-1,1:i-1}^{-1}$, and the inverse of a sparse triangular matrix is generally dense. For example, if $L_{ii}=1$, $L_{i,i-1}=a\neq0$ and all other entries are zero, then $(\bL^{-1})_{ij}=(-a)^{i-j}$ for all $j\le i$. Hence the conditional mean of $x_i$ generally depends on all of $\bx_{1:i-1}$, and the precision matrix $\bL^{-\top}\bL^{-1}$ of $\bx$ is generally dense. \qed

\subsection*{Proof of Proposition~\ref{prop:forecast}}

Given $\by_{1:t}$, $\bx_t\sim\normal(\bfnu_t,\bL_t\bL_t^\top)$, and $\bw_{t+1}$ is independent of $\bx_t$ and $\by_{1:t}$. Hence $\bx_{t+1}=\bH\bx_t+\bw_{t+1}$ is conditionally Gaussian with mean $\bH\bfnu_t$ and covariance $\bH\bL_t\bL_t^\top\bH^\top+\bQ=\bF\bF^\top+\bQ$, or equivalently, in MA form, $\bx_{t+1}=\bH\bfnu_t+\bF\bz+\bw_{t+1}$ with $\bz\sim\normal(\bfzero,\bI_n)$.

For (a), row $i$ of $\bF=\bH\bL_t$ is $\bF_{i,:}=\sum_kH_{ik}(\bL_t)_{k,:}$, a linear combination of the at most $c_H$ rows of $\bL_t$ with $H_{ik}\neq0$, each of which has at most $m_L$ nonzeros. The row therefore has at most $c_Hm_L$ nonzeros, and with a sparse accumulator it is computed in $\order(c_Hm_L)$ operations, or $\order(nc_Hm_L)$ for all $n$ rows. For (b), we have $[\bfSigma_{t+1|t}]_{kl}=\bF_{k,:}\bF_{l,:}^\top+Q_{kl}$, where the first term is an inner product of two sparse rows with at most $c_Hm_L$ nonzeros each, costing $\order(c_Hm_L)$, and the second term costs $\order(1)$ by assumption.

For (c), recall from \eqref{eq:vecchia} that column $i$ of $\hU_{t+1}$ is determined by the submatrix $[\bfSigma_{t+1|t}]_{\cK_i,\cK_i}$. The distances $\ell_i$ are nonincreasing along the maximin ordering, because the set of previously ordered locations grows. Hence, for $k\le l$ in $\cK_i$, we have $l\le i$ and $\|\bs_k-\bs_l\|\le\|\bs_k-\bs_i\|+\|\bs_i-\bs_l\|<2\rho\ell_i\le2\rho\ell_l$, so that $(k,l)\in\bS_{2\rho}$, and all required entries lie in $\bS_{2\rho}\cup\bS_{2\rho}^\top$. For each $l$, the locations $\bs_k$, $k<l$, are at least $\ell_l$ apart, because $\|\bs_k-\bs_{k'}\|\ge\ell_{k'}\ge\ell_l$ for $k<k'<l$. Comparing the volume of the disjoint balls of radius $\ell_l/2$ around those within distance $2\rho\ell_l$ of $\bs_l$ with that of the ball of radius $(2\rho+1/2)\ell_l$ around $\bs_l$ shows that there are at most $(4\rho+1)^d$ of them, so that $\bS_{2\rho}$ has $\order(n\rho^d)$ entries. Computing each required entry once, at cost $\order(c_Hm_L)$ by (b), takes $\order(n\rho^dc_Hm_L)$ time, computing $\bF$ takes $\order(nc_Hm_L)$ time by (a), and the local computations \eqref{eq:vecchia} take $\sum_i\order(|\cK_i|^3)=\order(nm_S^3)$ time. Only the sparse matrices $\bL_t$, $\bH$, and $\bF$ and the $\order(n\rho^d)$ required entries of $\bfSigma_{t+1|t}$ are stored, none of which has more than $\order(nc_Hm_L)$ nonzeros.

For (d), let $H_{kl}\neq0$, so that $\|\bs_k-\bs_l\|\le\rho\ell_n$. For $j\in\cJ_l$, we have $\|\bs_l-\bs_j\|<\rho\ell_j$, and hence $\|\bs_k-\bs_j\|<\rho\ell_n+\rho\ell_j\le2\rho\ell_j$, because $\ell_n\le\ell_j$. Row $k$ of $\bF$ is therefore supported on $\bar\cJ_k=\{j:\|\bs_k-\bs_j\|<2\rho\ell_j\}$. For quasi-uniform locations, the distances $\ell_j$, $j\ge2$, lie in $\order(\log n)$ dyadic intervals of the form $[h/2,h)$. The locations $\bs_j$ with $\ell_j\in[h/2,h)$ are at least $h/2$ apart, and those in $\bar\cJ_k$ lie within distance $2\rho h$ of $\bs_k$, so that, by the same volume argument, there are at most $(8\rho+1)^d$ of them. Hence $|\bar\cJ_k|=\order(\rho^d\log n)$, irrespective of $c_H$, and each entry in (c) costs $\order(\rho^d\log n)$. Moreover, all locations are at least $\ell_n$ apart, so that $c_H\le(2\rho+1)^d$, and $\bF$ is computed in $\order(n\rho^{2d}\log n)$ time. Together with the $\order(n\rho^d)$ entries of (c) and the local computations, the total time is $\order(n\rho^{3d}+n\rho^{2d}\log n)$. \qed

\subsection*{Proof of the error decomposition \eqref{eq:errdecomp} and of Proposition~\ref{prop:hv}}

We first prove the decomposition \eqref{eq:errdecomp} and its properties stated in Section~\ref{sec:hv}, and then use it to prove Proposition~\ref{prop:hv}. For symmetric positive definite $\bA$, \eqref{eq:kl} with $\bfSigma_P=\bA^{-1}$ gives
\[
\KL\big(Q\,\|\,\normal(\bfnu,\bA^{-1})\big)=\tfrac12\big[\operatorname{tr}(\bA\bfSigma_Q)-n-\log\det\bA-\log\det\bfSigma_Q\big].
\]
Subtracting the same expression with $\bB$ in place of $\bA$, the terms $-n$ and $\log\det\bfSigma_Q$ cancel, and since $\log\det\bA-\log\det\bB=\log\det(\bA\bB^{-1})$, we obtain
\begin{align*}
\KL\big(Q\|\normal(\bfnu,\bA^{-1})\big)-\KL\big(Q\|\normal(\bfnu,\bB^{-1})\big)
&=\tfrac12\big[\operatorname{tr}((\bA-\bB)\bfSigma_Q)-\log\det(\bA\bB^{-1})\big]\\
&=\Delta(\bA,\bB;\bfSigma_Q).
\end{align*}
Writing $\KL(Q\|P_{\mathrm{post}})=\KL(Q\|P_{\bV})+[\KL(Q\|\widetilde P)-\KL(Q\|P_{\bV})]+[\KL(Q\|P_{\mathrm{post}})-\KL(Q\|\widetilde P)]$ and applying this identity with $(\bA,\bB)=(\tW,\bV\bV^\top)$ and $(\bA,\bB)=(\bW,\tW)$ yields \eqref{eq:errdecomp}.

The properties of the decomposition follow directly. Since $\bW-\tW=(\bfSigma^{-1}+\bD^{-1})-(\hU\hU^\top+\bD^{-1})=\bfSigma^{-1}-\hU\hU^\top$, the noise precision $\bD$ cancels from the Stage-1 term. If Stage~1 is exact, then $\tW=\bW$, and if Stage~2 is exact, then $\bV\bV^\top=\tW$; in both cases, the corresponding term vanishes, because $\Delta(\bA,\bA;\cdot)=0$. If Stage~3 is exact, then $\bfSigma_Q=(\bV\bV^\top)^{-1}$ and $\KL(Q\|P_{\bV})=0$. Finally, for $\bfSigma_Q=\tW^{-1}$, using $\operatorname{tr}(\tW\tW^{-1})=n$ and \eqref{eq:kl} with $\bfSigma_Q=\tW^{-1}$ and $\bfSigma_P^{-1}=\bW$, we have
\[
\Delta(\bW,\tW;\tW^{-1})=\tfrac12\big[\operatorname{tr}(\bW\tW^{-1})-n-\log\det(\bW\tW^{-1})\big]=\KL(\widetilde P\|P_{\mathrm{post}}).
\]

We now turn to Proposition~\ref{prop:hv}, for which we need three auxiliary results. First, the HV prior factor is the Vecchia approximation with conditioning sets $\mathcal C_i$ (Section~\ref{app:hvsets}), and by \citet{Schafer2020}, the forward-KL-optimal factor on a given pattern is the Vecchia factor whose conditioning sets are given by that pattern. Applied to $\bS_{\mathrm{HV}}$, Stage~1 therefore returns $\hU_{\mathrm{HV}}$.

Second, incomplete Cholesky is exact whenever the exact factor has no fill-in outside the pattern. To see this, let $\bA\succ0$ and let $\bS$ be an upper-triangular pattern containing the diagonal with $\operatorname{supp}(\operatorname{rchol}(\bA))\subseteq\bS$. The exact factor is computed by a column-by-column recursion (e.g., Algorithm~1 of \citealp{Jurek2020}, applied after reversing the ordering), in which each entry is a function of $\bA$ and previously computed entries, and incomplete Cholesky uses the same recursion for the entries in $\bS$ while setting all other entries to zero. We argue by induction over the order of computation. If all previously computed entries of the two factors agree, then an entry in $\bS$ is computed from identical inputs and therefore agrees, whereas an entry outside $\bS$ is zero in the incomplete factor by construction and in the exact factor by assumption. Hence $\operatorname{ichol}(\bA,\bS)=\operatorname{rchol}(\bA)$.

Third, SPAI is exact whenever the exact inverse lies in the pattern. Indeed, if $\operatorname{supp}(\bV^{-\top})$ is contained in the Stage-3 pattern, then each row $\boldsymbol\ell_i=(\bV^{-\top})_{i,:}^\top$ is feasible and satisfies $\bV\boldsymbol\ell_i=\bV(\bV^{-1})_{:,i}=\mathbf e_i$, so its row residual $\|\mathbf e_i-\bV\boldsymbol\ell_i\|_2^2$ in the proof of Proposition~\ref{prop:klbound} is zero. Since zero is the smallest possible value and the minimizer of each row problem is unique, Stage~3 returns $\hL=\bV^{-\top}$.

With these results, (a)--(c) follow. For (a), by \citet[Prop.~2(1)]{Jurek2020}, $\operatorname{supp}(\operatorname{rchol}(\tW))\subseteq\bS_{\mathrm{HV}}$ under the hierarchical ordering, and the second result with $\bA=\tW$ and $\bS=\bS_{\mathrm{HV}}$ gives $\bV=\operatorname{rchol}(\tW)$. For (b), by \citet[Props.~1--2]{Jurek2020}, $\operatorname{chol}(\tW^{-1})$ has sparsity contained in $\bS_{\mathrm{HV}}^\top$. From (a), $\tW^{-1}=(\bV\bV^\top)^{-1}=\bV^{-\top}\bV^{-1}$, where $\bV^{-\top}$ is lower triangular with positive diagonal, so by uniqueness of the Cholesky factor, $\operatorname{chol}(\tW^{-1})=\bV^{-\top}$. Hence $\operatorname{supp}(\bV^{-\top})\subseteq\bS_{\mathrm{HV}}^\top$, and the third result gives $\hL=\bV^{-\top}$. For (c), (a) and (b) imply $\hL\hL^\top=\bV^{-\top}\bV^{-1}=\tW^{-1}$ with $\tW=\hU_{\mathrm{HV}}\hU_{\mathrm{HV}}^\top+\bD^{-1}$, so $Q_{\mathrm{HV}}=\widetilde P_{\mathrm{HV}}$. In \eqref{eq:errdecomp}, the Stage-3 and Stage-2 terms therefore vanish, and the Stage-1 term equals $\Delta(\bW,\tW;\tW^{-1})=\KL(\widetilde P_{\mathrm{HV}}\|P_{\mathrm{post}})$. \qed

\subsection*{Proof of the bounds in Remark~\ref{rem:bounds}}

For the first bound, let $\mathcal O=\{i:d_i<\infty\}$ be the set of observed locations, and let $\bt\mid\bx\sim\normal(\bx_{\mathcal O},\bD_{\mathcal O})$ be the (pseudo-)data, where $\bD_{\mathcal O}$ is the corresponding submatrix of $\bD$. Let $p$ and $q$ denote the joint distributions of $(\bx,\bt)$ under the priors $\bx\sim\normal(\bfmu,\bfSigma)$ and $\bx\sim\normal(\bfmu,(\hU\hU^\top)^{-1})$, respectively, with the same conditional distribution of $\bt$ given $\bx$. By the chain rule for the KL divergence,
\[
\KL\big(q(\bx)\,\|\,p(\bx)\big)=\KL\big(q(\bx,\bt)\,\|\,p(\bx,\bt)\big)=\KL\big(q(\bt)\,\|\,p(\bt)\big)+\bbE_{q(\bt)}\KL\big(q(\bx\mid\bt)\,\|\,p(\bx\mid\bt)\big),
\]
where the first equality holds because the conditional distributions of $\bt$ given $\bx$ agree. The posteriors are $q(\bx\mid\bt)=\normal(\bfnu_q(\bt),\tW^{-1})$ and $p(\bx\mid\bt)=\normal(\bfnu_p(\bt),\bW^{-1})$, whose means depend on $\bt$, but whose precision matrices do not. By the KL divergence between two Gaussian distributions,
\begin{align*}
\KL\big(q(\bx\mid\bt)\,\|\,p(\bx\mid\bt)\big)&=\KL(\widetilde P\|P_{\mathrm{post}})+\tfrac12\big(\bfnu_q(\bt)-\bfnu_p(\bt)\big)^\top\bW\big(\bfnu_q(\bt)-\bfnu_p(\bt)\big)\\
&\ge\KL(\widetilde P\|P_{\mathrm{post}})
\end{align*}
for every $\bt$. Here, $\widetilde P$ and $P_{\mathrm{post}}$ are the distributions of Section~\ref{sec:hv} of the main text, with the common mean $\bfnu$. Since $\KL(q(\bt)\|p(\bt))\ge0$, it follows that
\[
\KL(\widetilde P\|P_{\mathrm{post}})\le\KL\big(q(\bx)\|p(\bx)\big),
\]
which is the first bound. The expression for the Stage-1 term at a general $Q$ follows from the definition of $\Delta$, because its log-determinant term does not depend on $\bfSigma_Q$.

For the second bound, let $\lambda_1,\dots,\lambda_n>0$ be the eigenvalues of $\bfSigma^{1/2}\hU\hU^\top\bfSigma^{1/2}$, and let $g(\lambda)=\lambda-1-\log\lambda$. Then
\begin{align*}
\KL\big(\normal(\bfzero,\bfSigma)\|\normal(\bfzero,(\hU\hU^\top)^{-1})\big)&=\tfrac12\textstyle\sum_kg(\lambda_k),\\
\KL\big(\normal(\bfzero,(\hU\hU^\top)^{-1})\|\normal(\bfzero,\bfSigma)\big)&=\tfrac12\textstyle\sum_kg(1/\lambda_k).
\end{align*}
If the former is at most $\epsilon\le0.1$, then $g(\lambda_k)\le2\epsilon\le0.2$ for all $k$, which implies $\lambda_k\ge0.49$, because $g$ is decreasing on $(0,1]$ with $g(0.49)>0.2$. For $\lambda\ge1$, we have $g(1/\lambda)\le g(\lambda)$, because $g(\lambda)-g(1/\lambda)=\lambda-1/\lambda-2\log\lambda$ vanishes at $\lambda=1$ and has derivative $(1-1/\lambda)^2\ge0$, and an elementary computation shows that $g(1/\lambda)\le1.7\,g(\lambda)$ for $0.49\le\lambda<1$. Hence $\tfrac12\sum_kg(1/\lambda_k)\le1.7\cdot\tfrac12\sum_kg(\lambda_k)\le1.7\epsilon\le2\epsilon$. \qed

\section{Reverse-KL Optimal Sparse Covariance Factor\label{app:reversekl}}

This section shows the claim of Section~\ref{sec:stage3} of the main text that minimizing the reverse KL divergence over $\mathcal Q_{\mathrm{cov}}$ has a closed-form solution that decouples over the columns of the factor, and that the columns of coarse-scale locations require large solves. Let the target be $P=\normal(\bfnu,\tW^{-1})$, and let $Q_\bB=\normal(\bfnu,\bB\bB^\top)$ with $\bB\in\mathcal M(\SA)$ and positive diagonal. For column $j$ of $\SA$, let $\mathcal I_j=\{i\ge j:(i,j)\in\SA\}$, which contains $j$ as its smallest element, and let $\bb_j=\bB_{\mathcal I_j,j}$ denote the nonzero entries of column $j$ of $\bB$, so that $B_{jj}$ is the first entry of $\bb_j$. By \eqref{eq:kl} of the main text, and because $\bB$ is triangular,
\begin{align*}
\KL(Q_\bB\|P)&=\tfrac12\Big[\operatorname{tr}(\tW\bB\bB^\top)-n-\log\det\tW-2\textstyle\sum_{j=1}^n\log B_{jj}\Big]\\
&=\textstyle\sum_{j=1}^n\Big(\tfrac12\bb_j^\top\tW_{\mathcal I_j,\mathcal I_j}\bb_j-\log B_{jj}\Big)+\text{const},
\end{align*}
since $\operatorname{tr}(\tW\bB\bB^\top)=\sum_j\bB_{:,j}^\top\tW\bB_{:,j}$. The objective is a sum of strictly convex functions of the individual columns, which are minimized separately. Setting the gradient of the $j$th term to zero gives $\tW_{\mathcal I_j,\mathcal I_j}\bb_j=\mathbf e_1/B_{jj}$, where $\mathbf e_1$ is the first unit vector, so that $\bb_j=\tW_{\mathcal I_j,\mathcal I_j}^{-1}\mathbf e_1/B_{jj}$. Taking the first entry gives $B_{jj}^2=\mathbf e_1^\top\tW_{\mathcal I_j,\mathcal I_j}^{-1}\mathbf e_1$, and hence
\[
\widehat\bB_{\mathcal I_j,j}=\frac{\tW_{\mathcal I_j,\mathcal I_j}^{-1}\mathbf e_1}{\big(\mathbf e_1^\top\tW_{\mathcal I_j,\mathcal I_j}^{-1}\mathbf e_1\big)^{1/2}},
\]
the analogue of \eqref{eq:vecchia} of the main text for a covariance factor \citep[cf.][]{Schafer2020}. Column $j$ thus requires a solve with the $|\mathcal I_j|\times|\mathcal I_j|$ submatrix of $\tW$. The set $\mathcal I_j$ consists of the later locations within distance $\rho\ell_j$ of $\bs_j$ (Section~\ref{sec:patterns} of the main text). For the first location, $\ell_1=\infty$, so that $\mathcal I_1=\{1,\dots,n\}$ and the first column requires a solve with $\tW$ itself, and in general, the columns of coarse-scale locations have $\order(n)$ entries. In contrast, Stage~3 decouples over the rows, each of which involves only a $|\cJ_i|\times|\cJ_i|$ Gram matrix with $|\cJ_i|\le m_L$.

\section{Conditioning Sets of Hierarchical Vecchia\label{app:hvsets}}

We recall the conditioning sets of HV from \citet[Sec.~3.1 and eq.~(4)]{Jurek2020}. The domain is partitioned recursively: at level $0$, the region is $\domain$ itself, and each region at level $k-1$ is split into $J$ subregions at level $k$, $k=1,\dots,M$, giving the regions $\domain_{j_1,\dots,j_k}$, $j_1,\dots,j_k\in\{1,\dots,J\}$. Each region at level $k<M$ contains $r_k$ knots, chosen among its locations that are not knots of a coarser region, and at the finest level $M$, all remaining locations are knots. The locations are ordered by level, from coarse to fine. If $\bs_i$ is a knot of $\domain_{j_1,\dots,j_k}$, its conditioning set consists of all knots of its ancestor regions $\domain,\domain_{j_1},\dots,\domain_{j_1,\dots,j_{k-1}}$ and of the knots of its own region that are ordered before it,
\[
\mathcal C_i=\Big(\textstyle\bigcup_{l=0}^{k-1}\mathcal K_{j_1,\dots,j_l}\Big)\cup\{\bs_{i'}\in\mathcal K_{j_1,\dots,j_k}:i'<i\},
\]
where $\mathcal K_{j_1,\dots,j_l}$ denotes the knots of $\domain_{j_1,\dots,j_l}$. Hence $|\mathcal C_i|\le\sum_{l=0}^{k}r_l-1$, and the conditioning-set size $\cs$ of HV, as defined for all methods in Section~\ref{sec:setup} of the main text, is the median of $|\mathcal C_i|$ over $i$. Because $\mathcal C_i$ does not contain $\bs_i$, $\cs$ does not count $\bs_i$, as for \NNV{}, for which $\cs$ is the median of $|\cK_i|-1$. The sets are nested: if $\bs_{i'}\in\mathcal C_i$, then $\mathcal C_{i'}\subset\mathcal C_i$, because the region of $\bs_{i'}$ is $\domain_{j_1,\dots,j_k}$ or one of its ancestors, whose ancestors are ancestors of $\domain_{j_1,\dots,j_k}$. This closure is the reason why the exact Cholesky factors in Proposition~\ref{prop:hv} of the main text have no fill-in outside $\bS_{\mathrm{HV}}$ \citep[Props.~1--2]{Jurek2020}, and it allows our implementation of Section~\ref{sec:time} of the main text to compute Stages~2 and~3 of HV exactly within the pattern. The low-rank approximation (LR) is the special case $M=1$, in which every non-knot location conditions on the same $r_0$ knots. {Hence $\cs=r_0$ for LR.} In the GPvecchia implementation used in Section~\ref{sec:numerical} of the main text, $J=2$, and the numbers of knots are determined from the requested conditioning-set size.

\section{Details of the Basis Functions in Figure~\ref{fig:basis_functions}\label{app:basis}}

The $n=1{,}025$ locations are equally spaced on $[0,1]$, and the prior covariance is Mat\'ern with variance $1$, range $0.2$ and smoothness $1.5$. For \NNV, the locations are in maximin ordering, and the prior covariance factor is $\bL_{\text{prior}}=\hL$ obtained by applying Stage~3 with the pattern $\SA$ to the Stage-1 factor $\hU$ (i.e., with $\bV=\hU$). The resolution level of location $\bs_i$ is defined by its distance $\ell_i$ to the nearest previously ordered location, as $\max\{0,\operatorname{round}(\log_2(\ell_2/\ell_i))\}$, with level $0$ for $\bs_1$ ($\ell_1=\infty$), so that $\ell_i$ halves from one level to the next. This gives $3$, $2$, $4$ and $8$ basis functions at levels $0$--$3$. For HV, the regions at level $k$ are the $2^k$ intervals $[j2^{-k},(j+1)2^{-k})$, $j=0,\dots,2^k-1$ ($J=2$), the knots of a region are its first $r_k$ locations in the maximin ordering that are not knots of a coarser region, and the finest level contains all remaining locations once no region has more than $r_k$ of them. We use $r_0=3$ and $r_1=r_2=r_3=1$, so that HV has the same number of basis functions as \NNV{} at each of the four plotted levels, and $r_k=2$ at finer levels, as in \citet{Jurek2020}. This gives a maximum conditioning-set size of $16$, and the prior covariance factor of HV is $\bL_{\mathrm{HV}}=\hU_{\mathrm{HV}}^{-\top}$, which is the result of Stage~3 for the hierarchical pattern (Proposition~\ref{prop:hv} of the main text). For \NNV, $\rho=8.05$ is the smallest multiple of $0.05$ for which the maximum conditioning-set size $\max_i|\cK_i|-1$ reaches $16$. The figure shows all columns of $\bL_{\text{prior}}$ and $\bL_{\mathrm{HV}}$ at levels $0$--$3$.

\section{Details of the Numerical Experiments}\label{app:numdetails}

\subsection*{Methods and settings}

We use five sparsity levels with about $40$, $60$, $80$, $100$ and $120$ nonzeros per row. On the $40\times40$ grid of Sections~\ref{sec:sim_spatial} and~\ref{sec:sim_linear} of the main text, they are obtained with $\rho\in\{2.1,2.8,3.5,4.0,4.6\}$ for \NNV, which gives average conditioning-set sizes {$\sum_i(|\cK_i|-1)/n$} between $7$ and $37$ (maxima from $12$ to $68$); with requested GPvecchia conditioning-set sizes of $58$, $87$, $124$, $150$ and $181$ for HV (maximum conditioning-set sizes between $49$ and $153$); and with $\cs{}\in\{41,60,82,99,122\}$ for LR. The matched numbers of nonzeros per row agree to within $5\%$. \NNV{}~+ IC0 (Section~\ref{sec:setup} of the main text) replaces Stage~3 by the exact $\bV^{-\top}$. Its posterior mean is that of \NNV, and its log density only requires $\bV$, as $\sum_i\log V_{ii}-\|\bV^\top(\bx-\bfnu)\|^2/2$ up to a constant, so that its log score can be computed for any $n$; the exact $\bV^{-\top}$, which is needed for the forecast step, is only feasible for moderate $n$. For Gaussian data, the expectation of dLS over $\bx^{\mathrm{true}}$ and $\by$ equals the expected KL divergence from the exact to the approximate posterior. For non-Gaussian data, the exact method is itself a Laplace approximation, and the expectation equals $\KL(p_{\mathrm{true}}\|q)-\KL(p_{\mathrm{true}}\|p_{\mathrm{exact}})$, averaged over $\by$, where $p_{\mathrm{true}}$ is the true posterior; dLS then measures the loss of the approximation relative to the exact Laplace approximation. The Laplace iterations of \NNV{} converged at all time points, for all data sets and data models, within the iteration limit of Section~\ref{sec:stage1} of the main text. On average, they needed $1$ iteration for Gaussian data, $5$ for logistic data and about $8$ for Poisson and gamma data.

\subsection*{Large-$n$ experiments}

We compare the computation times and accuracy of the methods for $n$ up to $90{,}000$, with all methods matched at about $80$ nonzeros per row of the posterior covariance factor (Section~\ref{sec:setup} of the main text). For the spatial-only problem, we repeat the design of Section~\ref{sec:sim_spatial} of the main text on regular grids of $100\times100$, $200\times200$ and $300\times300$ locations ($n=10{,}000$, $40{,}000$ and $90{,}000$), with $10$ simulated data sets for each grid and data model. The settings are $\rho=2.6$, $2.23$ and $2.1$ for \NNV{} (average conditioning-set sizes {$\sum_i(|\cK_i|-1)/n$ of} $12.3$, $7.7$ and $7.4$), requested GPvecchia conditioning-set sizes of $105$, $102$ and $99$ for HV (maximum conditioning-set sizes $94$, $88$ and $88$), and $\cs{}=80$, $77$ and $78$ for LR. For filtering, we use the high-dimensional advection--diffusion setting of \citet[][Sec.~6.4]{Jurek2020} on the $300\times300$ grid: diffusion and advection coefficients $\alpha=10^{-7}$ and $\beta=10^{-3}$, $\bfSigma_{1|0}=\bQ_t$ Mat\'ern with variance $0.2$, range $0.15$ and smoothness $0.5$, a random $10\%$ of the locations ($n_t=9{,}000$) observed at each of $T=10$ time points, and Gaussian noise variance $\tau^2=0.05$, with $10$ simulated data sets.

All methods run in the CPU implementation of Section~\ref{sec:filter} of the main text. The times include Stage~1, all Newton iterations, and Stages~2 and~3, i.e., the computation of the posterior covariance factor needed for the forecast step, and for filtering also the forecast step. The ordering and the sparsity patterns, which depend only on the locations, are excluded for all methods. The exact posterior is only feasible at $n=10{,}000$. Following Section~\ref{sec:setup} of the main text, the log score $-\log q(\bx^{\mathrm{true}}\mid\by)$ of a method differs from its dLS only by a term that is the same for all methods, so that differences in log score between methods are differences in dLS. In the spatial-only comparison, \NNV{} had the best log score and the smallest RMSPE in every data set, and in filtering it had the smallest RMSPE at every time point of every data set and the best log score in all but $2$ of these $300$ cases. We therefore report the log score of HV and LR minus that of \NNV{}, and their RMSPE in percent above that of \NNV{}.

\section{Additional Numerical Experiments}\label{app:addexp}

\subsection{Boundary Effects of Hierarchical Vecchia}\label{sec:boundary}

\begin{figure}[htbp]
    \centering
    \includegraphics[width=0.78\textwidth]{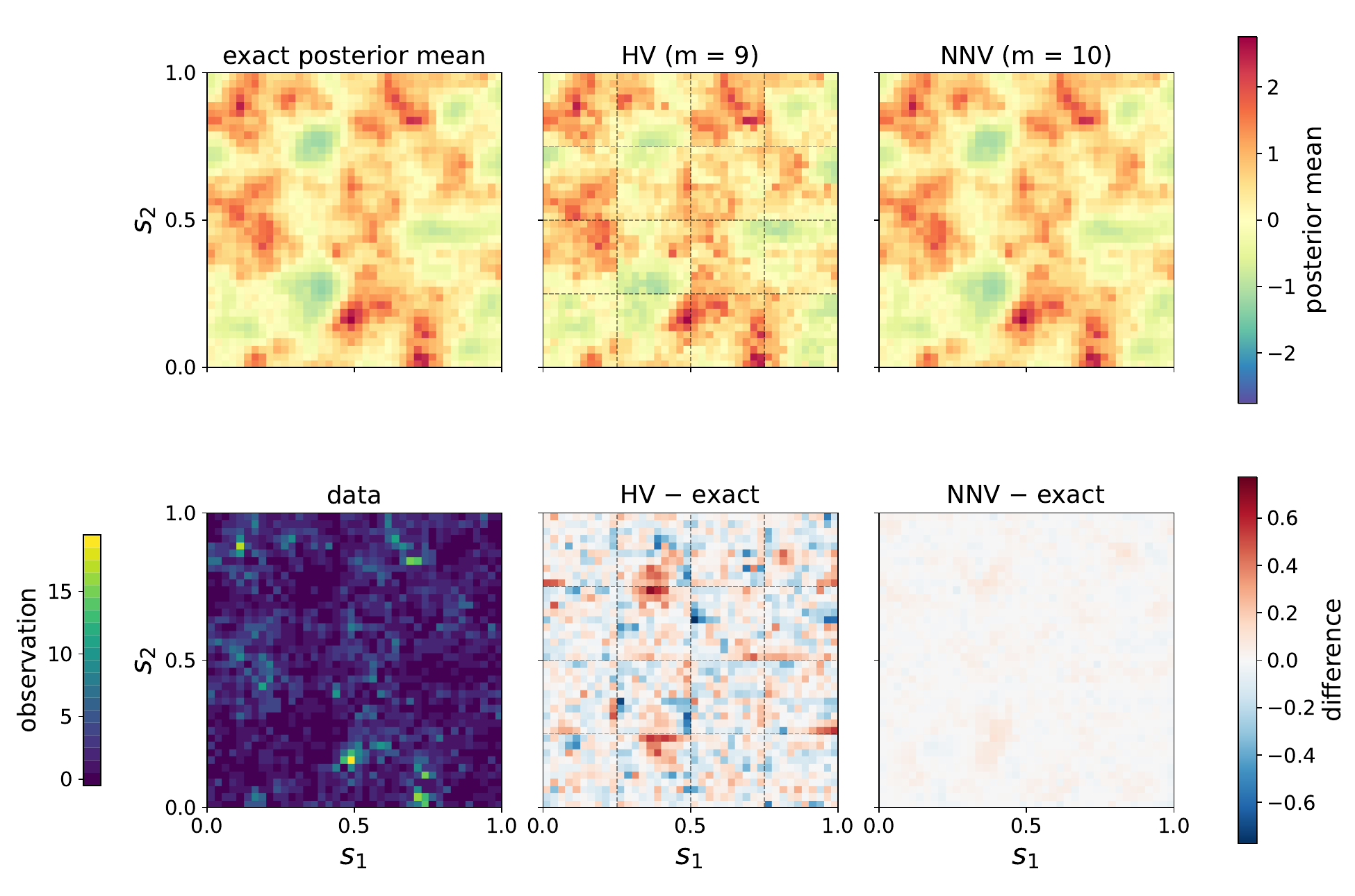}
    \caption{Simulated Poisson counts at $n=1{,}600$ locations with a log link (bottom left; Section~\ref{sec:boundary}). Top row: posterior means of the latent field using the exact Laplace approximation, HV ($\cs=9$), and \NNV{} ($\cs=10$), on a common color scale. Bottom center and right: differences of the HV and \NNV{} posterior means from the exact one, on a common color scale. Dashed lines in the HV panels mark the coarse partition lines of HV at $s_1,s_2\in\{1/4,1/2,3/4\}$.}
    \label{fig:boundary_poisson}
\end{figure}

The cut-off basis functions of HV have a direct consequence for the posterior means, as Figure~\ref{fig:boundary} of the main text (logistic data) and Figure~\ref{fig:boundary_poisson} (Poisson data) show. For these figures, we take the first simulated data set of Section~\ref{sec:sim_spatial} of the main text for the logistic and the Poisson data models, for which the observations carry the least information about the latent field, and compute the posterior means of the exact Laplace approximation, of HV and of \NNV{}. In GPvecchia, a requested conditioning-set size of $10$ gives $\cs{}=9$ for HV (mean $8.4$, maximum $9$), and for \NNV{}, $\rho=2.24$ is chosen by bisection such that the average conditioning-set size $\sum_i(|\cK_i|-1)/n$ is at least the requested size of $10$, which gives $\cs=10$ (mean $11.1$, maximum $20$). For $n=1{,}600$ and a requested size of $10$, GPvecchia sets $M=9$, with $J=2$ for the first eight levels and $J=8$ for the last one, and $r=1$ knot per region at levels $0,\dots,8$. This gives $511$ knots, and each of the remaining $1{,}089$ locations is conditioned on the $9$ knots of its regions at these levels. The splits alternate between the two coordinates, starting with $s_2$, so that the partition lines of levels $1$ to $6$ lie at $s_1,s_2\in\{1/8,\dots,7/8\}$. Figure~\ref{fig:boundary} of the main text shows the observations, the three posterior means, and the differences of the two approximations from the exact posterior mean for logistic data, and Figure~\ref{fig:boundary_poisson} shows them for Poisson data.

The \NNV{} posterior means cannot be distinguished visually from the exact ones. Their differences from the exact posterior mean are small and spatially homogeneous, with a root mean squared difference of $0.028$ for logistic and $0.022$ for Poisson data. The HV posterior means are instead visibly blocky. For logistic data in particular, they consist of rectangular patches and have a clear discontinuity along the line $s_1=1/2$, one of the two coarsest partition lines. The differences of HV from the exact posterior mean are about seven times larger, with a root mean squared difference of $0.185$ for logistic and $0.168$ for Poisson data, and they form rectangular patches with straight edges along the partition lines. To quantify this, we compute the mean absolute jump of the difference from the exact posterior mean between horizontally or vertically adjacent grid points. For HV, the jumps across the partition lines at multiples of $1/8$ are on average $2.2$ (logistic) and $1.6$ (Poisson) times as large as those across the other grid lines, whereas for \NNV{} this ratio is $1.0$ for both data models, i.e., \NNV{} has no preferred lines. The same pattern, with ratios of $1.3$ and $1.5$, holds for the Gaussian and gamma data models.

These boundary effects are caused by the cut-off basis functions of HV in Figure~\ref{fig:basis_functions} of the main text. By Proposition~\ref{prop:hv} of the main text, each column of the HV posterior covariance factor $\hL_{\mathrm{HV}}$, i.e., each posterior basis function, is zero outside its region. For two neighboring grid points on opposite sides of a partition line, only the basis functions at levels coarser than that of the line, whose regions contain both points, are nonzero at both of them, and all basis functions at the level of the line and finer are cut off at it. The posterior dependence between the two points, and hence the extent to which the data on one side of the line inform the posterior mean on the other side, is thus carried by only a few coarse basis functions (one per region and level here), although most of the dependence between neighboring points is at fine scales; for the exponential covariance with range $0.15$, two points at the grid spacing $1/40$ have a correlation of about $0.85$. The posterior mean therefore cannot borrow strength across the line and jumps there, most visibly at the coarsest lines $s_2=1/2$ and $s_1=1/2$, which are crossed by the fewest basis functions. The effect is strongest for logistic and Poisson data, for which a single observation says little about the latent value at its location, so that the posterior mean relies more on the prior dependence between neighbors. The basis functions of \NNV{}, in contrast, are centered at their own locations and overlap across any line in the domain, so that the dependence between neighbors is represented at all resolutions, and the approximation errors show no spatial structure.

\subsection{Prediction Errors}\label{app:rmspe}

\begin{figure}[ht!]
    \centering
    \includegraphics[width=\textwidth]{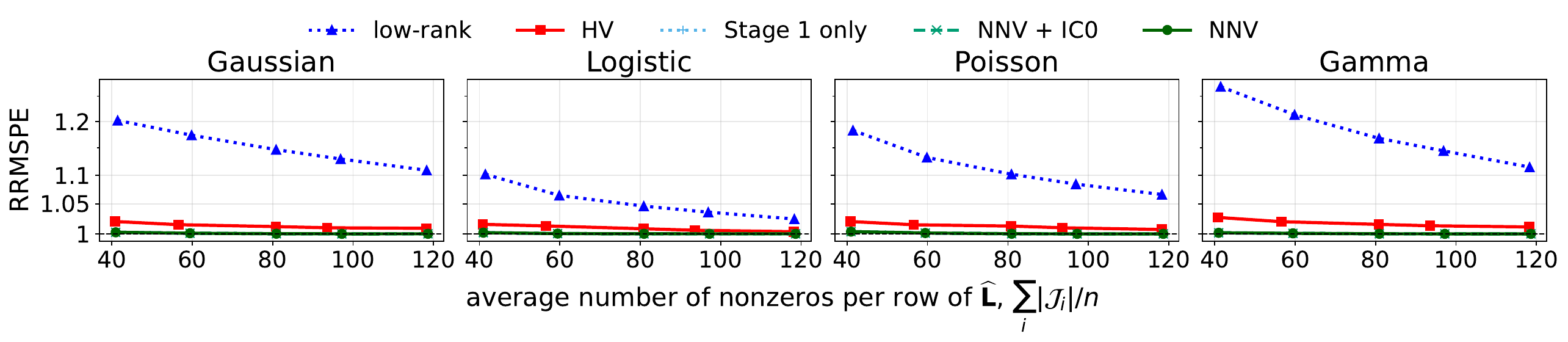}
    \caption{Spatial-only comparison ($n=1{,}600$; Section~\ref{sec:sim_spatial} of the main text): RRMSPE against the average number of nonzeros per row of the posterior covariance factor, for the methods and settings of Figure~\ref{fig:spatial} of the main text. The RRMSPE axis is on a log scale, and the dashed line marks RRMSPE${}=1$.}
    \label{fig:rmspe_spatial}
\end{figure}

\begin{figure}[ht!]
    \centering
    \includegraphics[width=\textwidth]{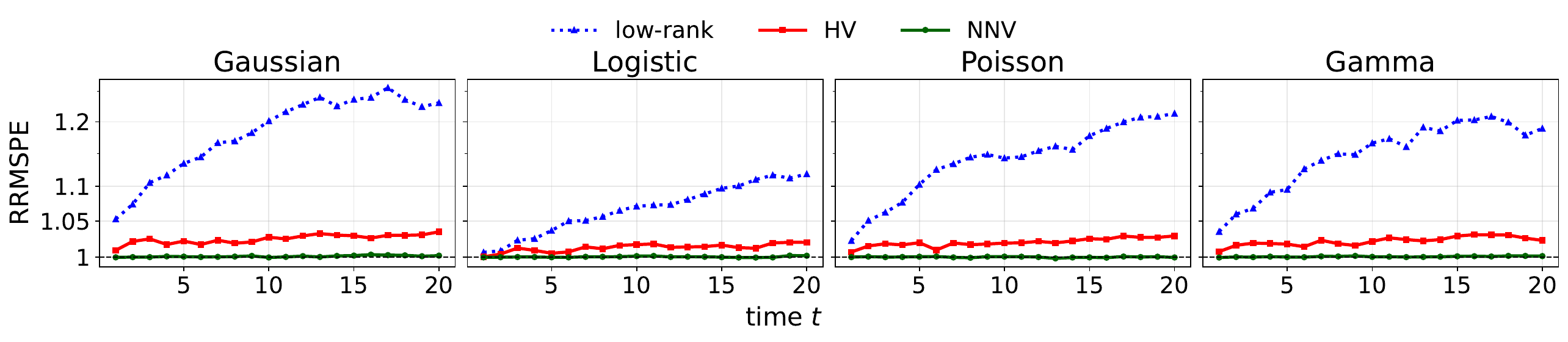}
    \caption{Linear advection--diffusion filtering ($n=1{,}600$; Section~\ref{sec:sim_linear} of the main text): RRMSPE of the filtering means over time, for the methods and settings of Figure~\ref{fig:linear} of the main text. The RRMSPE axis is on a log scale, and the dashed line marks RRMSPE${}=1$.}
    \label{fig:rmspe_linear}
\end{figure}

\begin{figure}[ht!]
    \centering
    \includegraphics[width=\textwidth]{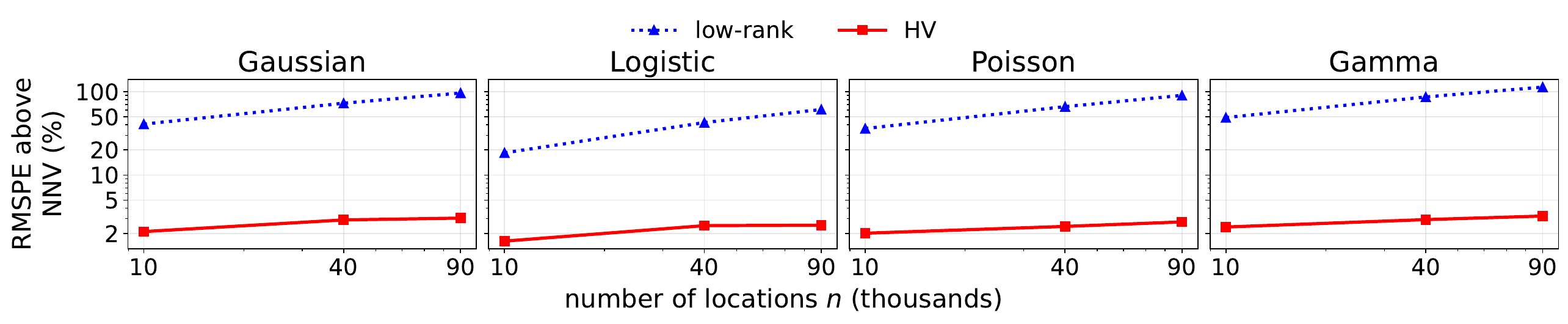}
    \caption{Spatial-only comparison for large $n$ (Section~\ref{sec:time} of the main text): RMSPE of HV and low-rank in percent above that of \NNV{}, against the number of locations $n$, for the settings of Figure~\ref{fig:time_spatial} of the main text. Both axes are on a log scale.}
    \label{fig:rmspe_large_n}
\end{figure}

Figures~\ref{fig:rmspe_spatial}--\ref{fig:rmspe_large_n} show the RMSPE of the posterior and filtering means for the comparisons in Figures~\ref{fig:spatial}, \ref{fig:linear} and~\ref{fig:time_spatial} of the main text.

\subsection{Filtering at All Sparsity Levels}\label{app:linear_sparsity}

\begin{figure}[htbp]
    \centering
    \includegraphics[width=\textwidth]{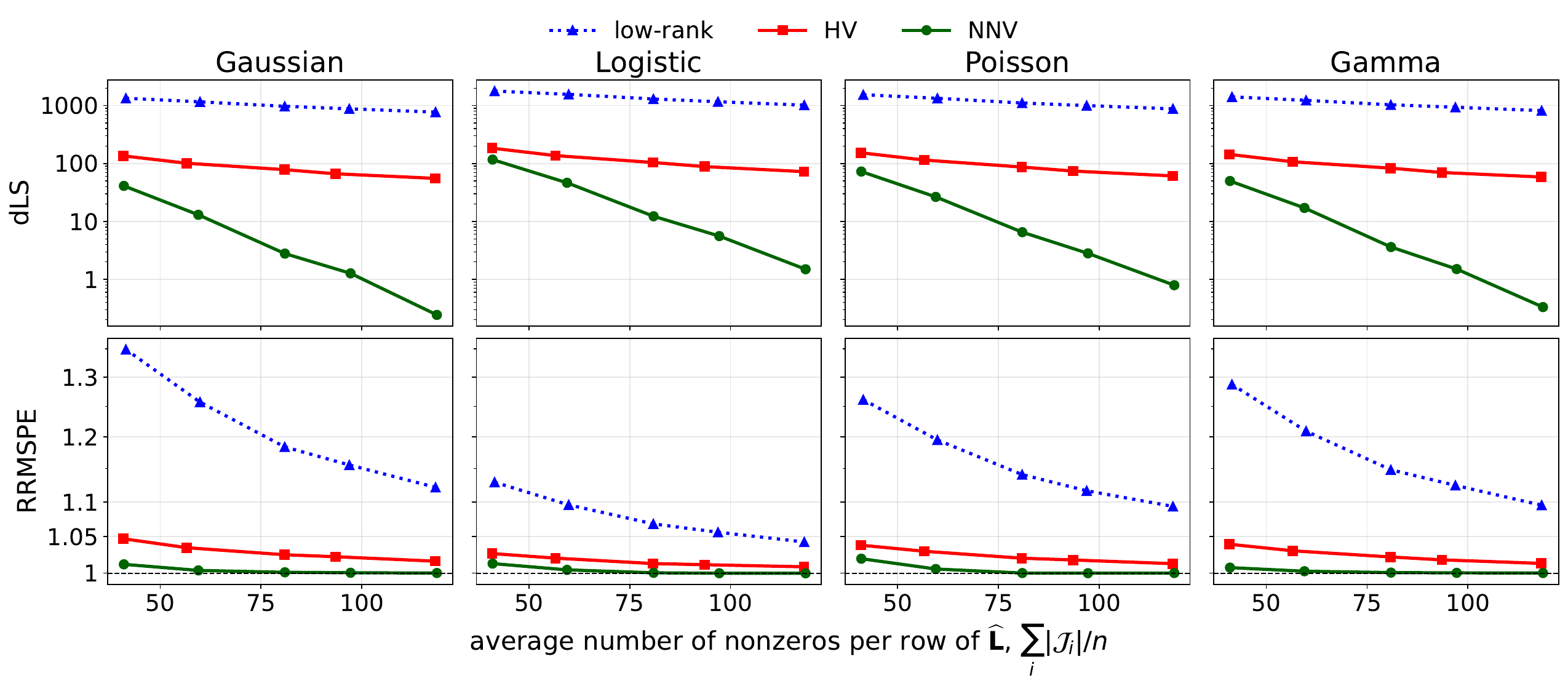}
    \caption{Linear advection--diffusion filtering ($n=1{,}600$; Section~\ref{sec:sim_linear} of the main text): dLS (top row) and RRMSPE (bottom row) of the filtering distributions, averaged over the $20$ time points and $10$ simulated data sets, against the average number of nonzeros per row of the covariance factor, $\sum_i|\cJ_i|/n$, for \NNV{}, HV and low-rank. Columns correspond to the Gaussian, logistic, Poisson and gamma data models. Both axes are on a log scale, and the dashed line marks RRMSPE${}=1$.}
    \label{fig:linear_sparsity}
\end{figure}

Figure~\ref{fig:linear_sparsity} shows the scores of the filtering experiment of Section~\ref{sec:sim_linear} of the main text, averaged over the $20$ time points, at all five sparsity levels of Section~\ref{sec:setup} of the main text.

\subsection{Filtering with Large $n$}\label{app:large_n_filter}

\begin{figure}[t!]
    \centering
    \includegraphics[width=0.78\textwidth]{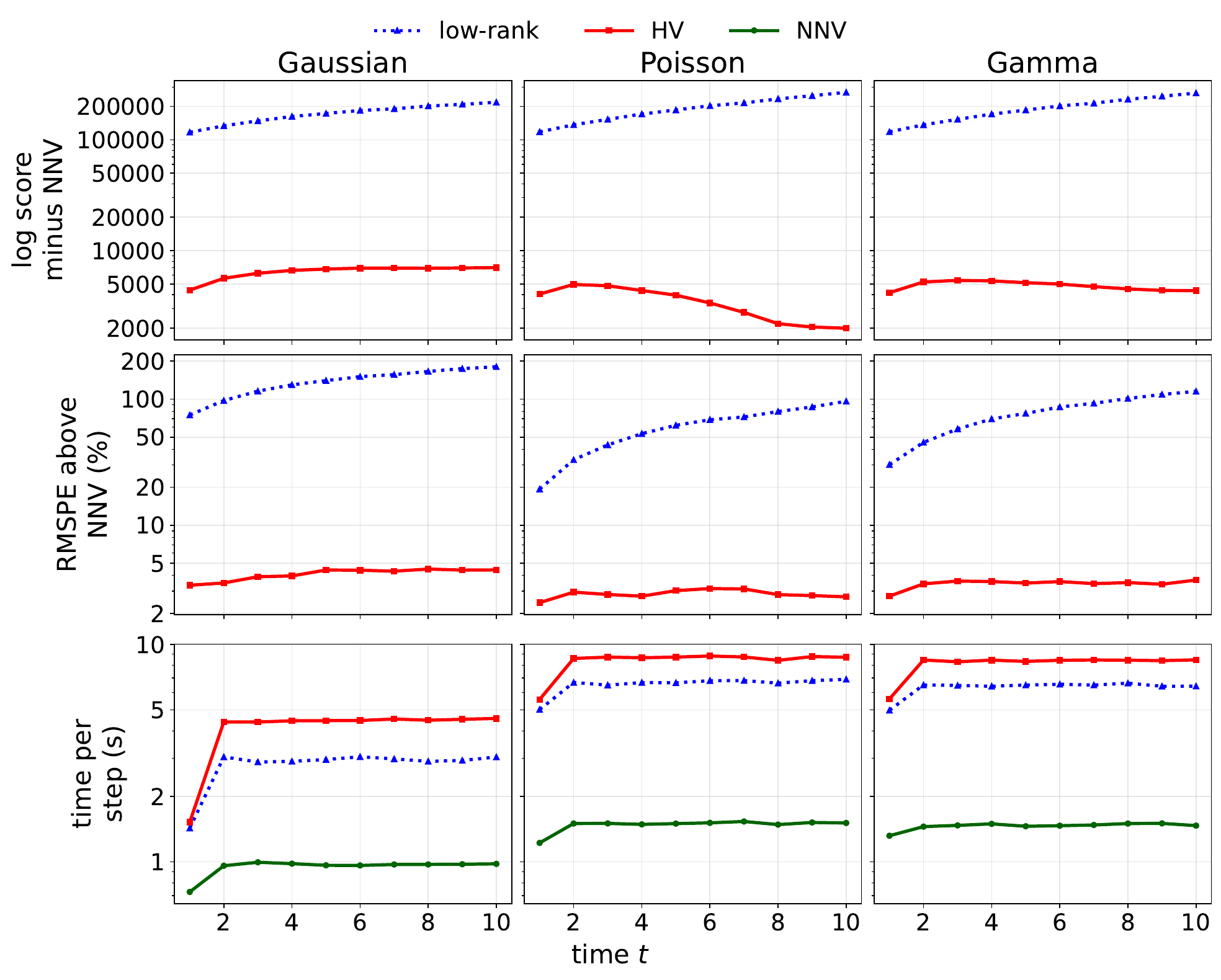}
    \caption{Filtering with $n=90{,}000$ (Section~\ref{sec:time} of the main text), with all methods at about $80$ nonzeros per row of the covariance factor: log score of HV and low-rank minus that of \NNV{} (top row), their RMSPE in percent above that of \NNV{} (middle row), and computation time per filtering step of all three methods (bottom row), as a function of time $t=1,\dots,10$. Columns correspond to the Gaussian, Poisson and gamma data models. Values are averaged over $10$ simulated data sets, and all axes are on a log scale.}
    \label{fig:time_filter}
\end{figure}

Figure~\ref{fig:time_filter} shows the corresponding results for filtering. A filtering step of \NNV{} is about four to six times as fast as one of HV and three to four times as fast as one of LR, and after the first time point, the time per step is nearly constant over time for all methods. At the same time, \NNV{} is more accurate than HV and LR at every time point, in log score and in RMSPE, for the three data models shown.

\FloatBarrier

\section{Additional Results of the Application}\label{app:application}

\begin{figure}[ht!]
    \centering
    \includegraphics[width=0.75\textwidth]{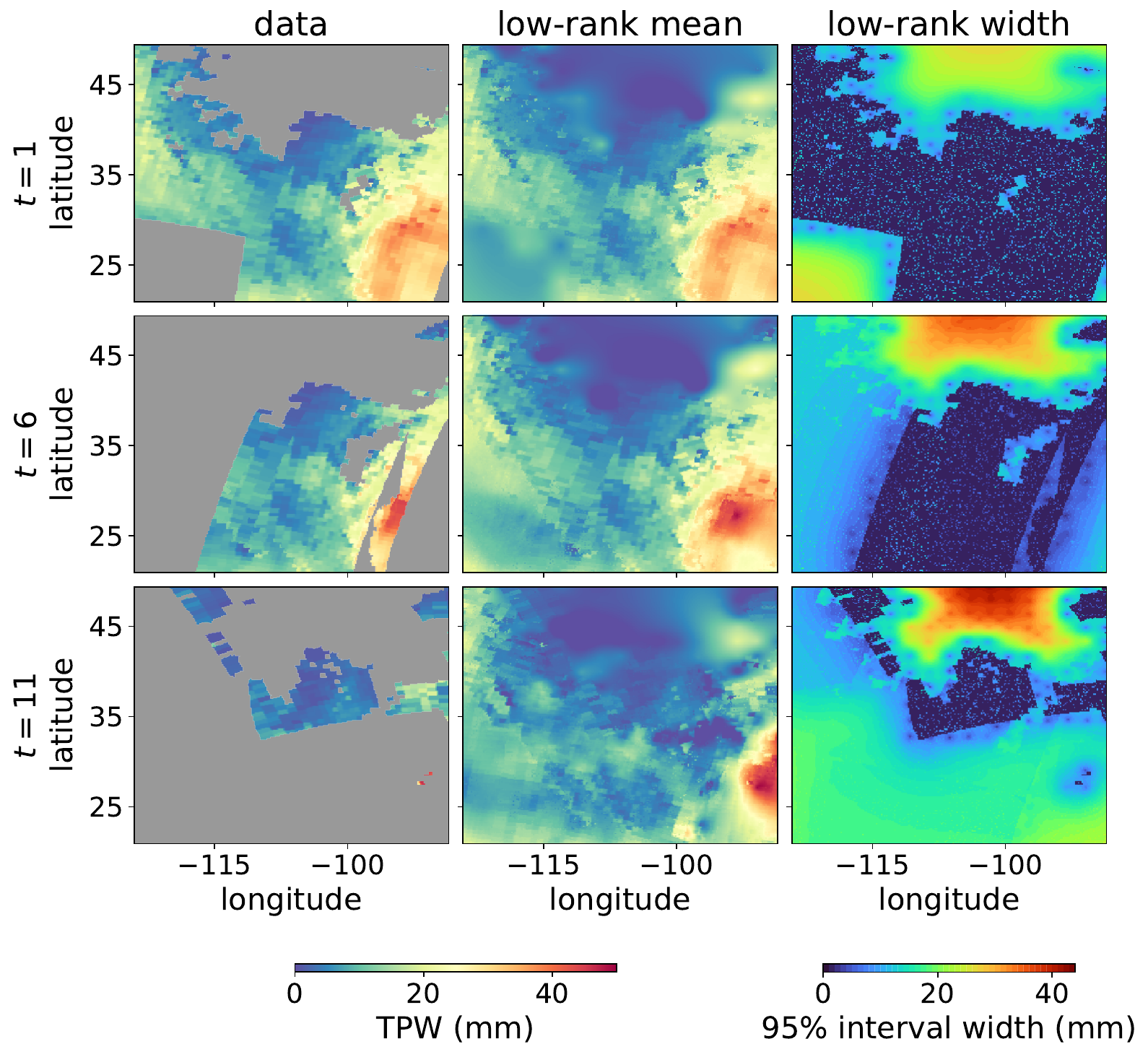}
    \caption{TPW application (Section~\ref{sec:application} of the main text) at $t=1$, $6$ and $11$ (rows): observations (including the held-out ones; gray cells are unobserved), filtering means and widths of the pixelwise $95\%$ prediction intervals of low-rank (discrete $1$-mm color bands, as in Figure~\ref{fig:application_maps} of the main text). The filter does not use the randomly held-out $10\%$ of the observations.}
    \label{fig:application_lowrank}
\end{figure}

Figure~\ref{fig:application_lowrank} shows the observations, the filtering means and the widths of the pixelwise $95\%$ prediction intervals of low-rank in the TPW application, as Figure~\ref{fig:application_maps} of the main text does for \NNV{} and HV. 

The size of the grid makes the exact Kalman filter just about feasible, which allows us to measure the approximation error of each method directly. With dense matrices, the exact filter stores the $n\times n$ filtering covariance matrix (about $30$\,GB) in memory and updates it in place. In the forecast step, $\bH_t$ has only five nonzeros per row, so that $\bH_t\bfSigma\bH_t^\top$ costs $\mathcal{O}(n^2)$, and $\bQ_t$ is added entrywise. In the update step, with $n_t$ observed cells, it computes the Cholesky factor of the $n_t\times n_t$ covariance matrix of the data, a triangular solve, and a rank-$n_t$ downdate of the covariance matrix at cost $\mathcal{O}(n^2n_t)$, using multithreaded LAPACK and BLAS routines on the same $32$ threads as the other methods (Section~\ref{sec:numerical} of the main text). These routines are highly optimized for dense matrices, whereas the approximations consist of many small and sparse operations in our Numba-compiled Python code, so that the comparison of computation times favors the exact filter. Because its cost depends on $n_t$, the time per step of the exact filter varies strongly with the number of observed cells, which ranges from $8\%$ to $60\%$ of the grid (Figure~\ref{fig:application_exact} of the main text), whereas that of \NNV{} is nearly constant. The memory of the exact filter grows quadratically in $n$ and, for a fixed fraction of observed cells, its cost per step cubically, so that on a grid with twice the resolution in each direction ($n\approx244{,}000$), it would need about $480$\,GB of memory and about $64$ times as long per step. The exact filter is therefore infeasible on substantially larger grids, especially for parameter estimation, which requires a full run of the filter for each of the many evaluations of the likelihood; even on this grid, one run takes about $37$ minutes (Table~\ref{tab:application} of the main text). 

\begin{figure}[ht!]
    \centering
    \includegraphics[width=\textwidth]{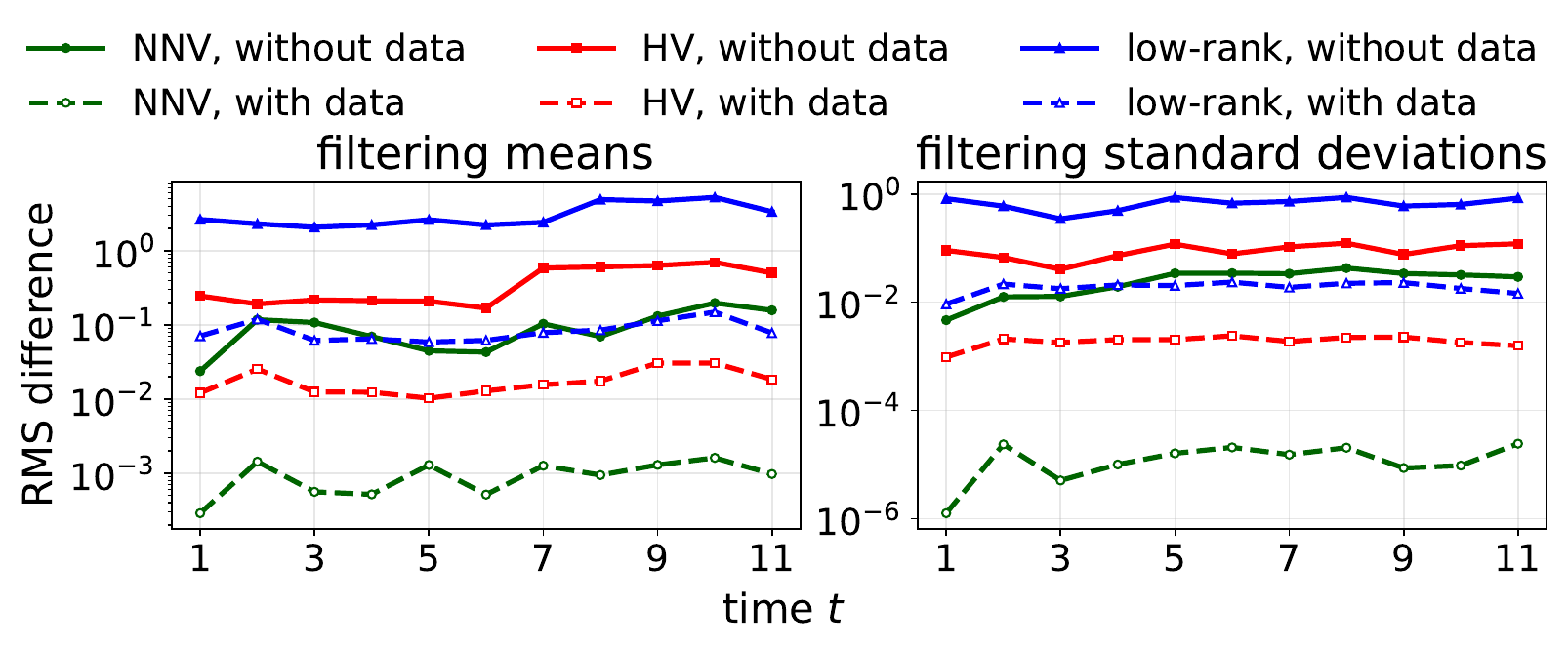}
    \caption{{TPW application (Section~\ref{sec:application} of the main text): root-mean-square differences (in mm) of the filtering means (left) and filtering standard deviations (right) of \NNV{}, HV and low-rank from those of the exact Kalman filter, over time, at the cells with training data at time $t$ (dashed lines, open markers) and at all other cells, including those with held-out data (solid lines, filled markers). Both axes are on a log scale.}}
    \label{fig:application_exact_rmsd}
\end{figure}

{Figure~\ref{fig:application_exact_rmsd} shows the root-mean-square differences of the filtering means and standard deviations of \NNV{}, HV and low-rank from those of the exact Kalman filter over time, separately for the cells with training data at time $t$ and for all other cells, including the cells with held-out data.}

\bibliographystyle{agsm}
\begingroup
\renewcommand{\bf}{\bfseries}
\renewcommand{\newline}{}
\def\harvardurl#1{}
\bibliography{mendeley,additionalrefs}
\endgroup